\documentclass[article]{jss}
\usepackage[toc,page]{appendix}

\author{Elyas Heidari\\Sharif Uni. of Tech. \And 
        Vahid Balazadeh-Meresht\\Sharif Uni. of Tech. \And
        Ali Sharifi-Zarchi\\Sharif Uni. of Tech.}
\title{Multivariate Analysis and Visualization using \proglang{R} Package \pkg{muvis}}
\Plaintitle{muvis: Multivariate Analysis and Visualization} 
\Shorttitle{\pkg{muvis}: Multivariate Analysis and Visualization} 
\usepackage{listings}
\usepackage{graphics}
\usepackage{subcaption}
\usepackage{float}
\usepackage{amsmath}
\usepackage{algorithm}
\usepackage{algorithmic}
\newlength{\fancyvrbtopsep}
\newlength{\fancyvrbpartopsep}
\makeatletter
\FV@AddToHook{\FV@ListParameterHook}{\topsep=\fancyvrbtopsep\partopsep=\fancyvrbpartopsep}
\makeatother
\usepackage{amsmath}

\usepackage{todonotes}
\usepackage{lscape}

\Abstract{
  Increased application of multivariate data in many scientific areas has considerably raised the complexity of analysis and interpretation. Although quite a few approaches have been suggested to address this issue, there is still a gap between the most efficient proposed methods and available software. \pkg{muvis} is an \proglang{R} package (\citet{Rcoreteam2017}) which is a toolkit for analyzing multivariate datasets. Several tools are implemented for common analyses of multivariate datasets, including preprocessing, dimensionality reduction, statistical analysis, Probabilistic Graphical Modeling, hypothesis testing, and visualization. Furthermore, we have implemented two novel methods--Variable-wise Kullback-Leibler Divergence (\code{VKL}) and Violating Variable-wise Kullback-Leibler Divergence (\code{VVKL})--in \pkg{muvis}, which are proposed to find the features with most different probability distributions between two groups of samples. The main aim of the package is to provide a wide range of users with different levels of expertise in \proglang{R} with a set of tools for comprehensive analysis of multivariate datasets. We exploited the NHANES dataset to declare the functionality of \pkg{muvis} in practice.
}

\Plainkeywords{Probabilistic Graphical Models, Variable-wise Kullback-Leibler Divergence, Multivariate Analysis, Multivariate Visualization, Statistical Modeling} 

\Address{
  Elyas Heidari\\
  Department of Computer Engineering,\\
  Sharif University of Technology\\
  Tehran, Iran\\
  E-mail: \email{eheidari@ce.sharif.edu}\\
  \\
  Vahid Balazadeh-meresht\\
  Department of Computer Engineering,\\
  Sharif University of Technology\\
  Tehran, Iran\\
  E-mail: \email{vbalazadehmeresht@ce.sharif.edu}\\
  \\
  Ali Sharifi-zarchi\\
  Department of Computer Engineering,\\
  Sharif University of Technology\\
  Tehran, Iran\\
  E-mail: \email{asharifi@sharif.ir}\\
}

\begin{document}

\section{Introduction}
The recent advent of huge data in a wide range of scientific fields such as sociology, environmental research, economics, and biomedical research has raised demands for methods and tools to interpret and analyze high-dimensional data, where each dataset contains a large number of measurements or variables. Quite a few approaches have been put forward to analyze multivariate datasets (\citet{Timm2004}, \citet{coghlan2014little}, \citet{esbensen2002multivariate}). There are a number of widely-used approaches for analysis of multivariate data, including hypothesis testing to assess the significance of the association between two variables, fitting linear or non-linear models to associate one feature to another feature or a set of other features in the dataset, and using correlation analysis to capture how variables interact with or influence each other (\citet{cohen2014applied}).

 The usual practice for analyzing multivariate data includes several steps: (i) Pre-processing and quality assurance, including identification and filtering of outliers and low-quality samples and missing data imputation. (ii) Multivariate analysis with possibly many different approaches, including dimensionality reduction, hypothesis testing, predictive models, correlation analysis, graphical modeling, etc. (iii) Visualization and interpretation of the results, including uni-, bi- and multivariate plots, interactive and dynamic graphical representations,  network visualization of interactions, etc. Several \proglang{R} packages have been developed so far for carrying out any of the above tasks (\citet{Tsagris2016}, \citet{Everitt2011}, \citet{Le2008}). Since there is no single package providing all of these functionalities, conducting the whole analysis requires using many different packages and re-adaptation of data among them, which is a cumbersome challenge for many people of different scientific backgrounds who need to analyze their multivariate data.
 
Here we introduce \pkg{muvis} as a comprehensive toolkit for multivariate analysis and visualization providing an end-to-end analysis pipeline. Furthermore, we highlight the necessity of the paradigm shift from regular correlation analysis to Probabilistic Graphical Modeling in multivariate settings. Additionally, we introduce two novel distribution-based methods based on Kullback-Leibler Divergence analogous to hypothesis testing (we use the term \textit{KL-based methods} to refer to these two methods). These methods will be introduced in details in the following sections.  

This paper is organized as follows: Section \ref{sec_thoery} gives a brief theoretical background of Graphical Models (GMs) and the novel KL-based methods. In section \ref{sec_implement}, the implementation of the package functions is described in details. Section \ref{sec_data} contains the results of the package, applied on a real multivariate dataset. Our conclusions are drawn in the final section. 

\section{Theoretical Background}
\label{sec_thoery}
\subsection{Graphical Models}
\label{graphical_models}
One of the key challenges in the analysis of a large dataset, containing many variables (i.e., measurements) and observations (i.e., samples), is to capture the associations among variables and represent them in a simple manner. Consider a dataset of $n$ variables measured in $m$ observations. One can investigate $\binom{n}{2}$ pairwise associations among variables, which is computationally intensive and difficult to interpret. Therefore, the objective is to prune the complete graph to one containing a subset of key associations rather than all possible links. To this end, one of the most acclaimed approaches is to use Probabilistic Graphical Models (PGMs), in which partial (conditional) dependencies among variables are represented as a sparse graph.

Graphical Models (GMs) are renowned for modeling relations among variables in a compact manner. Based on principles in probability and graph theory, they supply effective tools to deal with complexity as well as uncertainty underlying the structure of associations among variables. 
More precisely, given a multivariate random variable, PGMs are aimed to describe the probability distribution of the variable which is equivalent to the structure of the partial dependencies among variables (\citet{Lauritzen1996}, \citet{Koller2007}). The most distinctive feature of GMs is partial independence structure in the joint distribution of the variables which is often sparse even in complex phenomena. This leads to a sparse representation of the dependency structure. The theory behind GMs is described in more details in the following section. In parallel with theoretical developments, several software packages are developed for analysis of data using GMs. Particularly, \proglang{R} community has made a significant contribution in this regard (\citet{Højsgaard2012}). \\

\begin{itemize}
    \item \textbf{\large Markov Networks (Markov Random Fields):} A Markov Random Field is a joint probability distribution of a number of Gaussian random variables  $X_1, X_2, \dots, X_n$ represented as an undirected graph $G$.
    Each node of $G$ represents a variable, and each edge indicates a non-zero partial correlation between a pair of variables (\citet{Rue2005}). Here, we focus on Gaussian Graphical Models (GGMs). Gaussianity is proposed to be a reasonable assumption according to its mathematical simplicity and its dominance in nature (\citet{Veron2003}; \citet{uhler2017gaussian}). 
    
    \begin{itemize}
        \item \textit{Conditional Dependence:} In many statistical analyses, the problem is to find the relationships (dependence structure) among a subset of variables given occurrence of an event. This concept is defined as conditional dependence in statistics. Given three random variables, $X_i$, $X_j$, and $X_k$, $X_i$ and $X_j$ are independent conditioned on $X_k$ if and only if 
        \begin{equation}
            P(X_i;X_j|X_k) = P(X_i|X_k)P(X_j|X_k).
        \end{equation}

        \item \textit{Precision Matrix:} Let $X = (X_1, X_2, \dots, X_n)$ be an $n$-dimensional normally-distributed random vector. Assuming $X\sim N_n(\mu, \Sigma)$, the density function of $X$ can be shown as
        \begin{equation}
            f_{\mu,\Sigma}(x) = (2\pi)^{-n/2}(det\Sigma)^{-1/2}e^{-\frac{1}{2}(x-\mu)^T\Sigma^{-1}(x-\mu)},
        \end{equation}

were $\mu$ is mean and $\Sigma$ is the covariance matrix of X. We call $\Sigma^{-1}$ the concentration/precision matrix. An
interesting property of this formula is that the precision matrix consists of partial correlations such that $\Sigma^{-1}_{ij}$ is equal to the partial correlation of $X_i$ and $X_j$ (\citet{wasserman2013all}). It can be shown that zero partial correlation is equivalent to partial independence. Throughout this paper we will use zero (non-zero) partial correlation and partial independence (dependence) alternatively. 
Therefore, a normal distribution can be represented as a graph $G = (V, E)$ in which $V$ is the set of vertices (nodes), each node $V_i$ representing a variable $X_i$, and $E$ is the set of edges such that for an edge $e_{ij}$ exists if and only if $\Sigma^{-1}_{ij} \neq 0$. We call such graph a GGM. The objective is to estimate the precision matrix.
Given $m$ identically independent distributed observations $X^{(1)}, X^{(2)}, \dots, X^{(m)}$ from $N_n(\mu,\Sigma)$ the log-likelihood function can be written as \\

\begin{align}
l(\mu,\Sigma)& = C(-\frac{m}{2}log\, det(\Sigma) - \frac{1}{2}\Sigma_{i=1}^{m}(X^{(i)}-\mu)^T\Sigma^{-1}(X^{(i)}-\mu)) \nonumber\\
& \propto \frac{m}{2}log\,det(\Sigma)- \frac{1}{2}tr(\Sigma^{-1}(\Sigma_{i=1}^m(X^{(i)}-\mu)(X^{(i)}-\mu)^T)\\
&=-\frac{m}{2}log\,det(\Sigma)-\frac{m}{2}tr(S\Sigma^{-1})-\frac{m}{2}(\bar{X}-\mu)^T\Sigma^{-1}(\bar{X}-\mu),   \nonumber
\end{align}

where $C$ is a positive constant. In addition, given $\mu = \bar{X}$, the term $-\frac{n}{2}(\bar{X}-\mu)^T\Sigma^{-1}(\bar{X}-\mu)$ is constant. 
Thus, in order to solve Maximum Likelihood (ML) estimation, we have to maximize the term $-log\,det(\Sigma)-tr(S\Sigma^{-1})$ on all possible covariance matrices $\Sigma$ or if we assign $K = \Sigma^{-1}$ then we should maximize $log\,det(K)+tr(S\Sigma^{-1})$ for $K$ in the set of all possible precision matrices, where $S = \Sigma_{i=1}^n(X^{(i)}-\mu)(X^{(i)}-\mu)^T = \hat{\Sigma}$ (\citet{Lauritzen1996}). The aim now is to construct a GM, $G$.

    \item \textit{\large Model Selection:} So as to construct the GM, we must select the best model among all possible models, i.e., for a graph with $v$ nodes we have to select the most likely graph fitting the model among all graphs with $v$ nodes and possible $\binom{v}{2}$ edges.
    \begin{itemize}
        \item \textit{Stepwise Methods:} A collection of methods proposed for model selection are based on the stepwise approach. In this method, we start from a graph with no edge (or all possible edges) and follow sufficient forward (or backward) steps to construct the graph. We introduce two stepwise methods implemented in our package, below:
        \begin{itemize}
            \item \textit{Akaike Information Criterion (AIC):} Akaike Information Criterion is based on minimizing negative of log-likelihood, penalized by model complexity (\citet{Burnham2010}). The AIC of the model, with parameter $k$, is
            \begin{equation}
                AIC(k) = -2l + k|E|,
            \end{equation}

in which $|E|$, number of edges, stands for model complexity (or degree of freedom). In AIC, at each step of forward (or backward) stepwise algorithm, we add (or remove) the edge which minimizes the AIC (or decrease it over some threshold). 
            
            \item \textit{Bayesian Information Criterion (BIC):} Bayesian Information Criterion is very similar to AIC except for $k$ which has to be equal to $log(m)$ when we have $m$ observations of each variable (\citet{Claeskens2008}).
        \end{itemize}
        
        \item \textit{Thresholding:} A very simple method for constructing GGMs is to threshold partial correlations. In this method, the edge between two nodes with partial correlation above a threshold will be added to, otherwise will be eliminated from the graph. However, one should estimate the whole precision matrix prior to use this method and the estimation may not be feasible due to non-positive-definiteness of the covariance matrix.
        
        \item \textit{Significance Tests:} According to Gaussianity, one can test if a partial correlation is zero. To this end, consider the estimation of $\rho_{i,j}$ as the partial correlation between variables $i$ and $j$, it can be shown that one has
        \begin{equation}
            \hat{\rho}_{i,j} = S_{ij}-S_{i,V - \{i,j\}}S^{-1}_{V - \{i,j\},V - \{i,j\}}S_{V - \{i,j\},j}.
        \end{equation}

Now consider Fisher's z-transform as 
\begin{equation}
    \hat{z}_{i,j}=\frac{1}{2}log\big(\frac{1+\rho_{i,j}}{1-\rho_{i,j}}\big).
\end{equation}

The test statistic $T_m = \sqrt{m-p+2-3|\hat{z}_{i,j}|}$ can be used with a rejection of $R_n = (-F^{-1}(1-\alpha/2),F^{-1}(1-\alpha/2))$, with $F$, the cumulative distribution of standard normal, to a test of power $\alpha$ (\citet{Drton2007}). 

        \item \textit{glasso:} glasso proposed to maximize the penalized log-likelihood function,
        \
        \begin{equation}
            l-\lambda |K|_1
        \end{equation}

were $\lambda$ is penalizing parameter to determine model sparsity and $|K|_1$ is the sum of absolute values of off-diagonals of the precision matrix to control the model complexity. This leads to a convex programming problem which is straightforward to solve. Additionally, glasso algorithm can be applied well to high-dimensional settings (\citet{Friedman2008}).
    \end{itemize}
    \end{itemize}
    \item \textbf{\large Bayesian Networks (Causal Networks):} In every multivariate setting, an interesting investigation is to find causal effects among the variables, that is, to find which variable, measurement, or feature is a cause of another. Although yet there has been no algorithm proposed to capture the whole causal association set among a set of variables, there are some algorithms with satisfying efficiency, developed for causal inference (\citet{Drton2007}). The formalism of the problem is as follows:\\
Assume $X$ to be an $n$-dimensional random variable with density function $f$. One can factorize $f$ as
\begin{equation}
    f_X(x) = \prod_{i=1}^n f_{X_i}(x_i|Pa(X_i)),
\end{equation}

where $Pa(X_i) \subseteq \{X_1, X_2, \dots, X_n\}$ ($Pa$ stands for $Pa$rent). A Bayesian network is a directed graph like $G=(V,E)$, where $V$ is the vertex set in which each vertex, like $v_i$, represents a variable, like $X_i$, and each edge, like $e_{ij}$, in the set edge $E$ stands for conditional dependence between the two variables, like $X_i$ and $X_j$, given all other variables. In $G$ each edge ending at $v_i$ is either directed from corresponding vertices of $Pa(X_i)$ to $v_i$ or from $v_i$ to $v_j$ iff $X_i \in Pa(X_j)$. Accordingly, the direction of each edge implies probabilistic causality, since, given the states of parents, the probability distribution of child will be determined. It can be shown that this setting leads to a Directed Acyclic Graph (DAG) representing $f$ due to \textbf{Markov property} (\citet{Drton2007}). Thus, in order to construct a Bayesian network one must first estimate such a factorization of $f$. To that end, one approach is to first find conditional (partial) independence structure of the multivariate model and then find directions (causal relations) among the structure. In the next section, we introduce a constraint-based algorithm we implemented in \pkg{muvis}, named FCI.
    \begin{itemize}
        \item \textit{Fast Causal Inference Algorithm:} FCI is a constraint-based algorithm first proposed in (\citet{Kalisch2010a}). This class of algorithms, aim to find some constraints given observed data that are necessary if the variables have a specific causal structure. Afterward, the causal structure will be estimated according to the set of constraints in hand.
FCI is a generalization of \textbf{PC-algorithm} (\citet{Kalisch2010}). PC-algorithm consists of three prominent steps. In the beginning, the undirected \textbf{skeleton} of the graph is estimated. Then, for each edge like $(u,v)$ which is present in the graph, the algorithm checks if there is any subset of nodes that can separate the two ends of the edge $(u,v)$, i.e., if there is any subset like $S$ which $\text{u \& v are independent}|S$. This step can be carried out by checking the constraints mentioned before. In the next step, the $v1-v2-v3$ substructures (we will call such substructures, $v-structures$) are oriented with some rules (\citet{Kalisch2010}). Finally, the algorithms use some rules to orient further edges avoiding directed cycles (\citet{Kalisch2010}).
In FCI is based upon PC-algorithm assuming the existence of some hidden variables. The first part of the FCI algorithm is the same as the PC-algorithm. In light of the existence of hidden variables, excluding edges due to some conditional subset separations is not sufficient. So the algorithm uses more rules to remove more edges due to the possibility of the presence of hidden variables. Description of the details of these algorithms is beyond the scope of this paper (readers can find the details in \citet{Kalisch2010a} and \citet{kalisch2018overview}). There is also a faster version of FCI which is computationally cheaper and is known as an approximation of FCI, named RFCI.
    \end{itemize}
    \item \textbf{\large Minimal Forest for High-dimensional Modeling:} When the number of variables is too large (hundreds and thousands of variables), simple Graphical Modeling algorithms may fail both statistically (efficiency) and computationally (performance). Correspondingly, proper algorithms should be used in order to address these issues in high-dimensional settings. In the following, we will introduce an algorithm called minimal forest which is designed for high-dimensional Graphical Modeling.
    \begin{itemize}
        \item \textit{The Chow-Liu Algorithm:} Chow and Liu proposed an algorithm based on maximum weight spanning tree algorithm to find the maximum likelihood tree for multinomial discrete distributions. The algorithm is fast enough to be applied to high-dimensional data. The formulation comes in the following.

        Given an $m \times n$ dataset with $m$ observations of $n$ discrete variables, we aim to fit a maximum likelihood tree to the variables. Suppose that $V$ is the set of variables (nodes) and $E$ is the set of associations (edges). \citet{Chow1968} showed that the probability of observing $V = v$ can be written as
        
        \begin{equation}
            P(v) = \frac{\prod_{(V_i,V_j)\in E} P(V_i = v_i, V_j = v_j)}{\prod_{i \in V}P(V_i=v_i)^{d_i-1}}
        \end{equation}

        were $P$ is the probability distribution function and $d_i$ is the degree of the node $v_i$ in the tree. It can be shown that the maximum log-likelihood is the summation of mutual information between each pair of variables. Where the mutual information between $V_i$ and $V_j$ is defined as
        \begin{equation}
            I_{i,j} = \sum_{v_i, v_j} \sum I(V_i = v_i,V_j=v_j) log\frac{\sum I(V_i = v_i,V_j=v_j)}{\sum I(V_i = v_i) \sum I(V_j = v_j)},
        \end{equation}
        
        where $I$ is the indicator function. Thus, if we use $I_{i,j}$ as the weight of the edge $(V_i, V_j)$, applying the maximum spanning tree algorithm on the graph will lead to the maximum likelihood tree (\citet{Chow1968}). As mutual information can be also defined for continuous distributions, this method can be extended to continuous variables and also mixed distributions of continuous and discrete variables (\citet{Edwards2010}). Finally, so as to construct the maximum spanning tree from a connected graph, the Kruskal's algorithm can be used (\citet{Kruskal1956}).
        \item \textit{AIC/BIC minimal forest:} Although extracting maximum-likelihood spanning tree gives a spare representation of the associations within the set of variable, it will force the representation to be connected, which may not be true for some settings. Respectively, extending the tree to a forest will address this problem.
        \citet{Edwards2010} proposed a penalized mutual information measure based on AIC and BIC. In that paper they introduced $I^{AIC}_{i,j} = I_{i,j} - k|E|$ and $I^{BIC}_{i,j} = I_{i,j} - log(m)|E|$ as the alternative weights of the edge $(V_i,V_j)$. After filtering out the edges with negative weights the Kruskal's algorithm can be employed to select the maximum spanning tree of the graph (\citet{Edwards2010}). 
    \end{itemize}
\end{itemize}
\subsection{Variable-wise KL-divergence}

\textbf{\large Kullback-Leibler Divergence:} Kullback-Leibler divergence (KL-divergence) is a measure of dissimilarity of one probability distributions from another distribution (Fig. \ref{KL}a). Given $D_{\mathrm {KL}}(P||Q)$ as the KL-divergence of distribution $Q$ with respect to distribution $P$, $D_{\mathrm {KL}}(P||Q)$ indicates a measure of error, assuming $Q$ when the real distribution is $P$ (\citet{kullback1951information}). KL-divergence for discrete probability distributions $P$ and $Q$ is defined as
    \begin{equation}
        D_{\mathrm {KL} }(P\|Q)=\sum_{i}P(i)\,\log {\frac {P(i)}{Q(i)}}.
    \end{equation}
    
    Moreover, KL-divergence can be applied to continuous distributions as
    \begin{equation}
        {D_{\mathrm {KL} }(P\|Q)=\int _{-\infty }^{\infty }p(x)\,\log {\frac {p(x)}{q(x)}}\,dx}.
    \end{equation}
    
    It can be proved that KL-divergence is non-negative. One can use KL-divergence in a symmetric manner in order to find the distance between two distributions. Thus, $D_{\mathrm {sym}}(P,Q)$ may be defined as 
    \begin{equation}
        D_{\mathrm {sym}}(P,Q) = D_{\mathrm {KL} }(P\|Q) + D_{\mathrm {KL} }(Q\|P).
    \end{equation}
    \begin{itemize}
    \item \textbf{\large Variable-wise Kullback-Leibler Divergence:} The Symmetric KL-divergence which is defined in the previous section can be used to compute the divergence of two groups of observations of a single measurement. In other words, if the measurement can discriminate the two groups efficiently. One can generalize this to a set of variables (measurements) with two fixed groups of observations (Fig. \ref{KL}b). With such an approach, KL-divergence can be respected as a measure of importance along variables, thus, sorting variables regarding their KL-divergence values orders the list of variables in terms of importance. We call this approach Variable-wise KL-divergence (VKL).
    \item \textbf{\large Find Violating Variables with VKL:} Assuming two measurements (variables) to have a linear association, one can simplistically fit a line to the pair of variables. Outliers of a linear regression can be identified regarding their residuals. The observations with the most absolute values of residuals can be defined as outliers, by setting a cut-off. Subsequently, looking for differences (in measurements) of such outliers may lead to some valuable information (Fig. \ref{KL}c), i.e., these outliers should have some features which block (violate) the expected linear association. For instance, suppose that we have a dataset including some laboratory (continuous) measurements and a number of clinical demographic (discrete) data collected from a population and we want to look if there are some features that may violate the positive (linear) association of Body Mass Index (BMI) as a measure of obesity and Diastolic Blood Pressure (DBP). Accordingly, looking for differences between up-outliers (relative high BMI and low DBP) and low-outliers (relative low BMI and high DBP) in every other measurement we have will lead us to the most important features that are potential to violate the expected association. To this end, we can use VKL-divergence in order to find the most different features of the two groups of outliers. Such features may be very informative because of the ability to block the expected linear association. In this specific example, one may be interested to see which features can block high DBP in relatively obese individuals, and also which features can cause high DBP in relatively slim individuals (Fig. \ref{KL}c).
    \item \label{kl_pval} \textbf{\large Significance Levels of KL-divergence:} In order to compute the significance level of KL-divergence, one can permute the members between two groups of observations to see if the KL value for the variable is significant (size of the two groups should be fixed). In this setting, after creating a lot of permuted groupings, the true KL-divergence value can be compared to the empirical distribution constructed by permuting groups (Fig. \ref{KL_permute}). 
    \begin{figure}
    \centering
    \includegraphics[height=164mm, width=140mm]{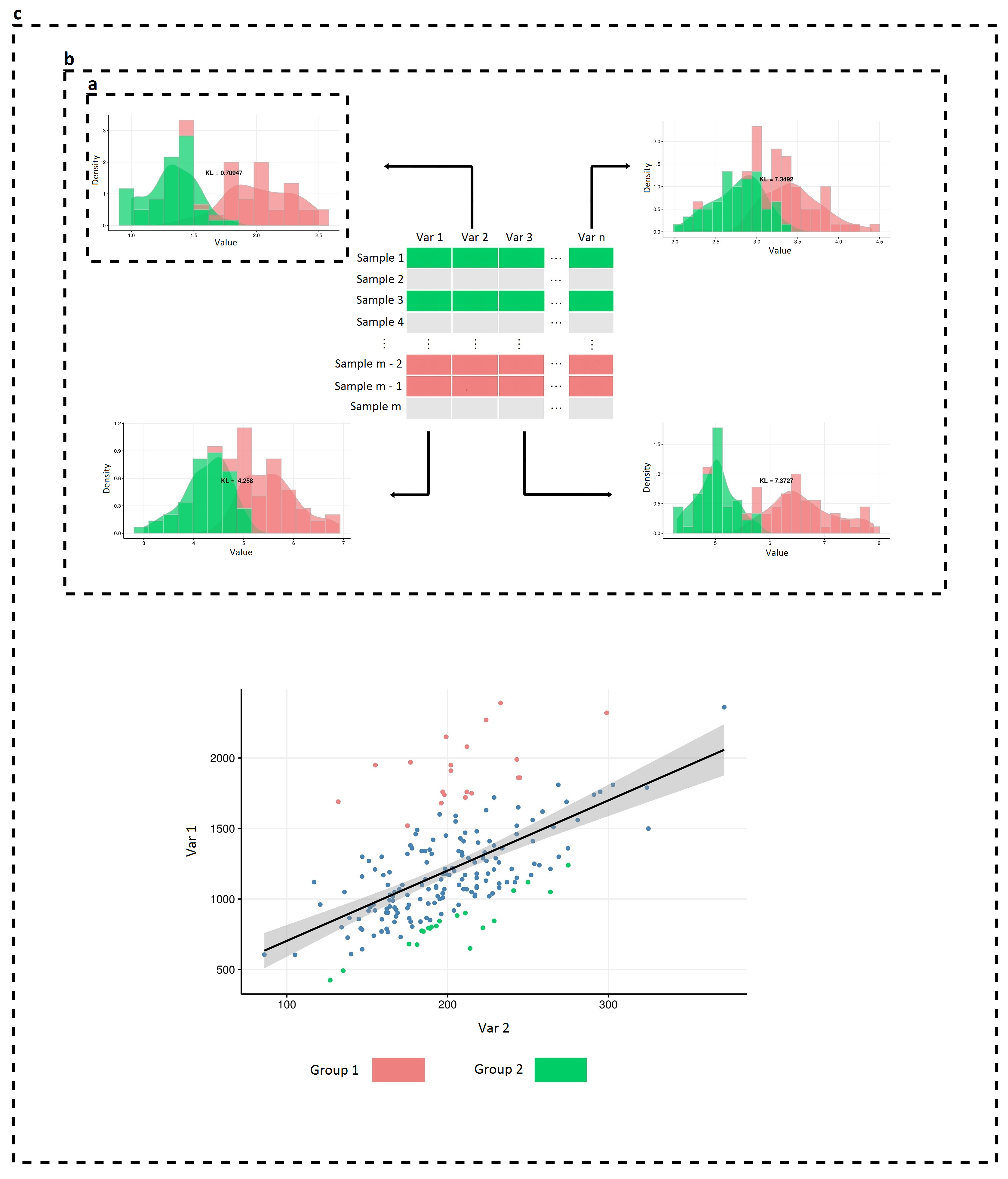}
    \caption{\textbf{Variable-wise KL-based methods.} Given a multivariate dataset with $m$ samples and $n$ variables, \textbf{a)} simple (symmetric) KL-divergence (KL) can be used in order to find the distance between probability distributions of two interesting groups of sample, colored in green and red, for a single variable; \textbf{b)} Variable-wise KL-divergence (VKL) can be used so as to calculate KL-divergence between two groups of samples for each variable, by calculating KL-divergence between the two groups on each variable; \textbf{c)} by fitting a linear model on two variables of interest, Violating Variable-wise KL-divergence (VVKL), finds the outliers with most absolute residuals in the linear model, the upper group is colored in red and the lower one is in green and these two groups of outliers will be then passed to VKL  to find the violating (blocking) variables for the linear association expected on the two variables.}
    \label{KL}
\end{figure}
\begin{figure}
    \centering
    \includegraphics[scale=1]{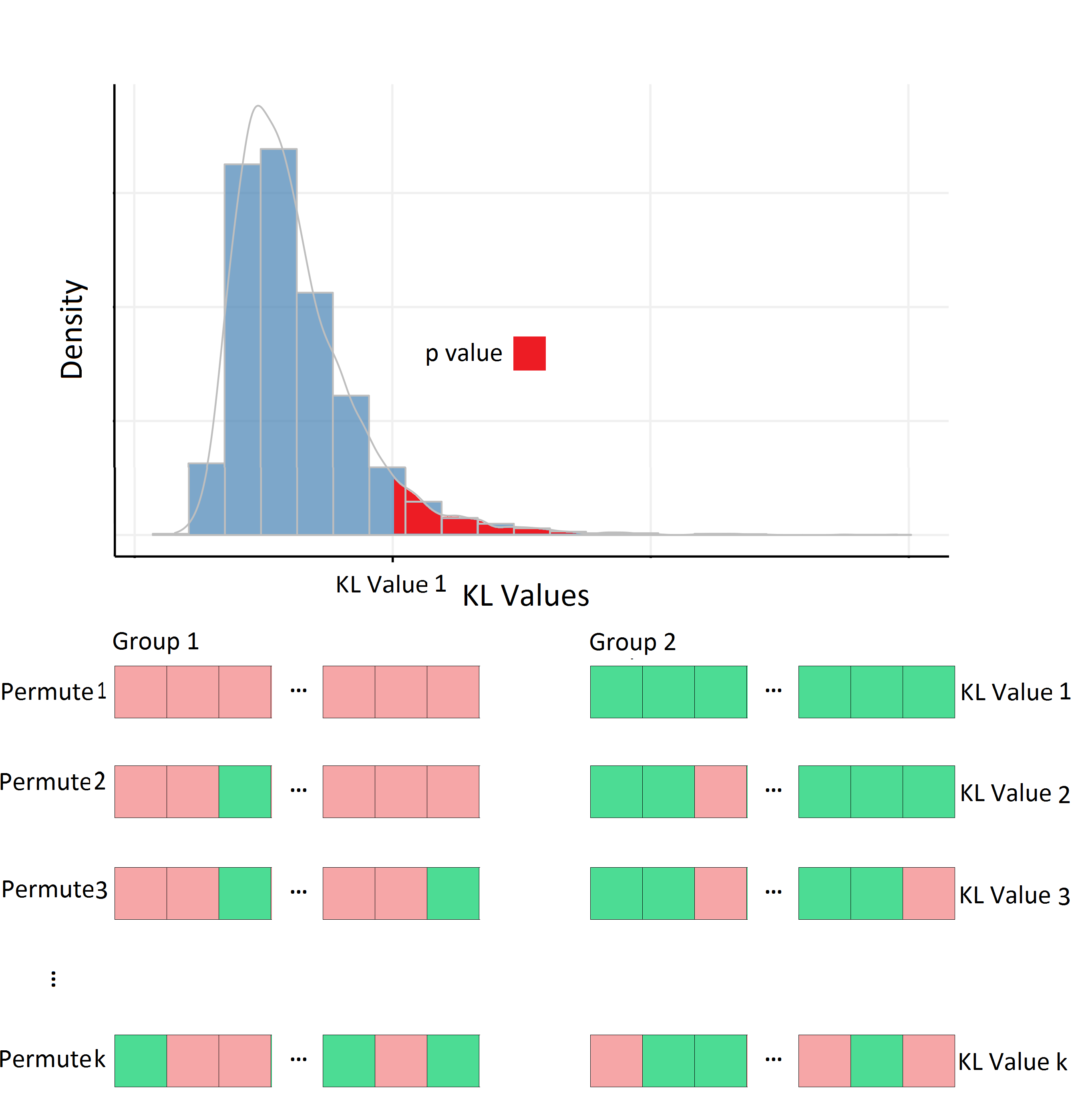}
    \caption{\textbf{The significance level of KL-divergence}. In order to find the significance level of a KL-divergence (KL Value 1) of a variable between two groups of samples (Permute 1), one can permute the samples between groups for $k$ times and compute the KL-divergence between new couple of groups (Permute 2, Permute 3, $\dots$, Permute k). Afterward, one can find the significance level of the KL-value by considering the empirical distribution of permuted KL-values as it is illustrated in the upper plot.}
    \label{KL_permute}
\end{figure}
\end{itemize}
\section{Package Implementation}
\label{sec_implement}
\subsection{Preprocessing}
\begin{figure}
    \centering
    \includegraphics[width = 150mm]{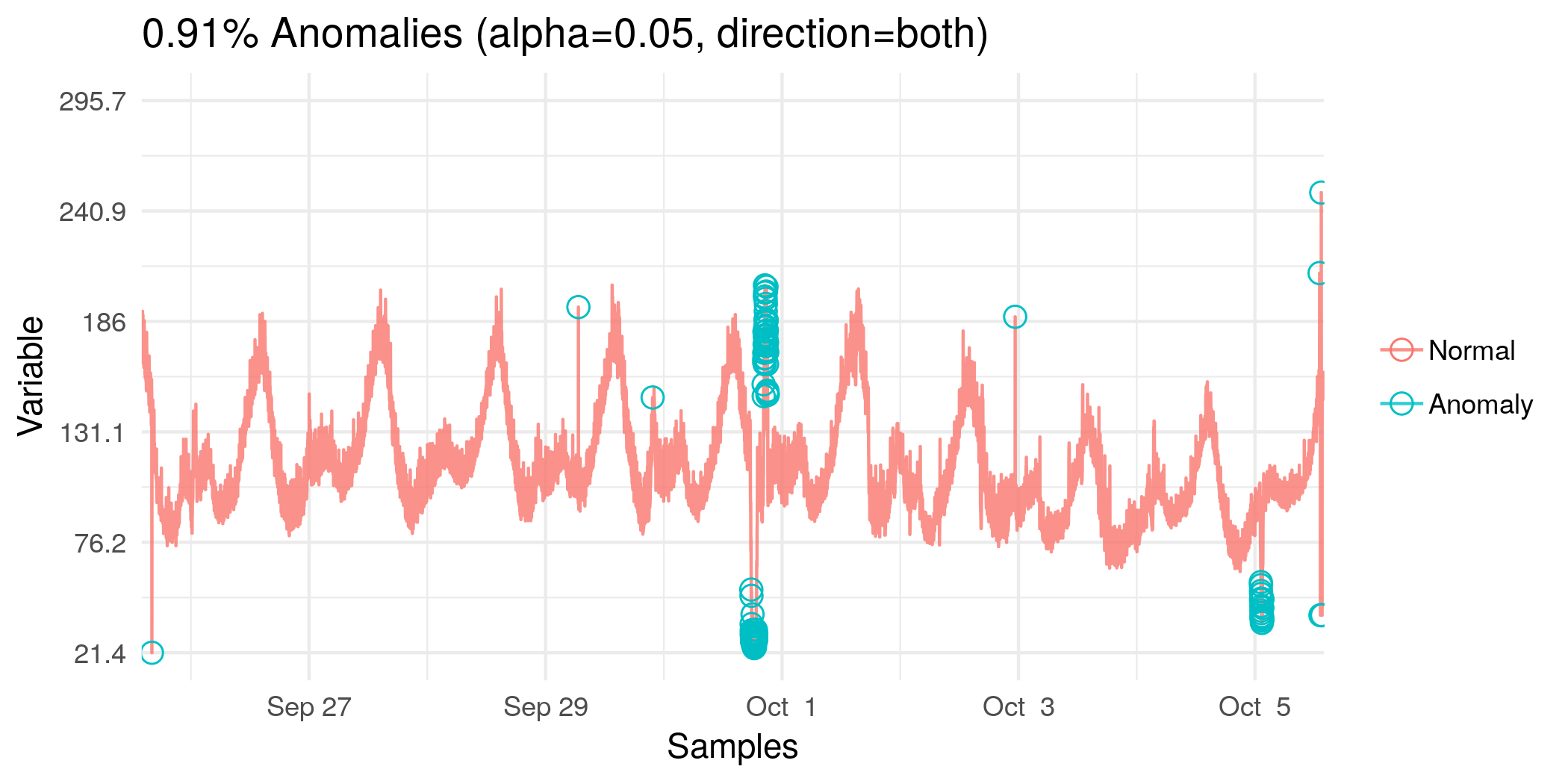}
    \caption{\textbf{Anomaly Detection by \citet{Vallis2014}}. The points in time-series data which do not follow the major trend of data are recognized as anomalies. The normal samples are colored in red and the anomalies are in blue}
    \label{anom}
\end{figure}

\begin{figure}
    \centering
    \includegraphics[width = 150mm]{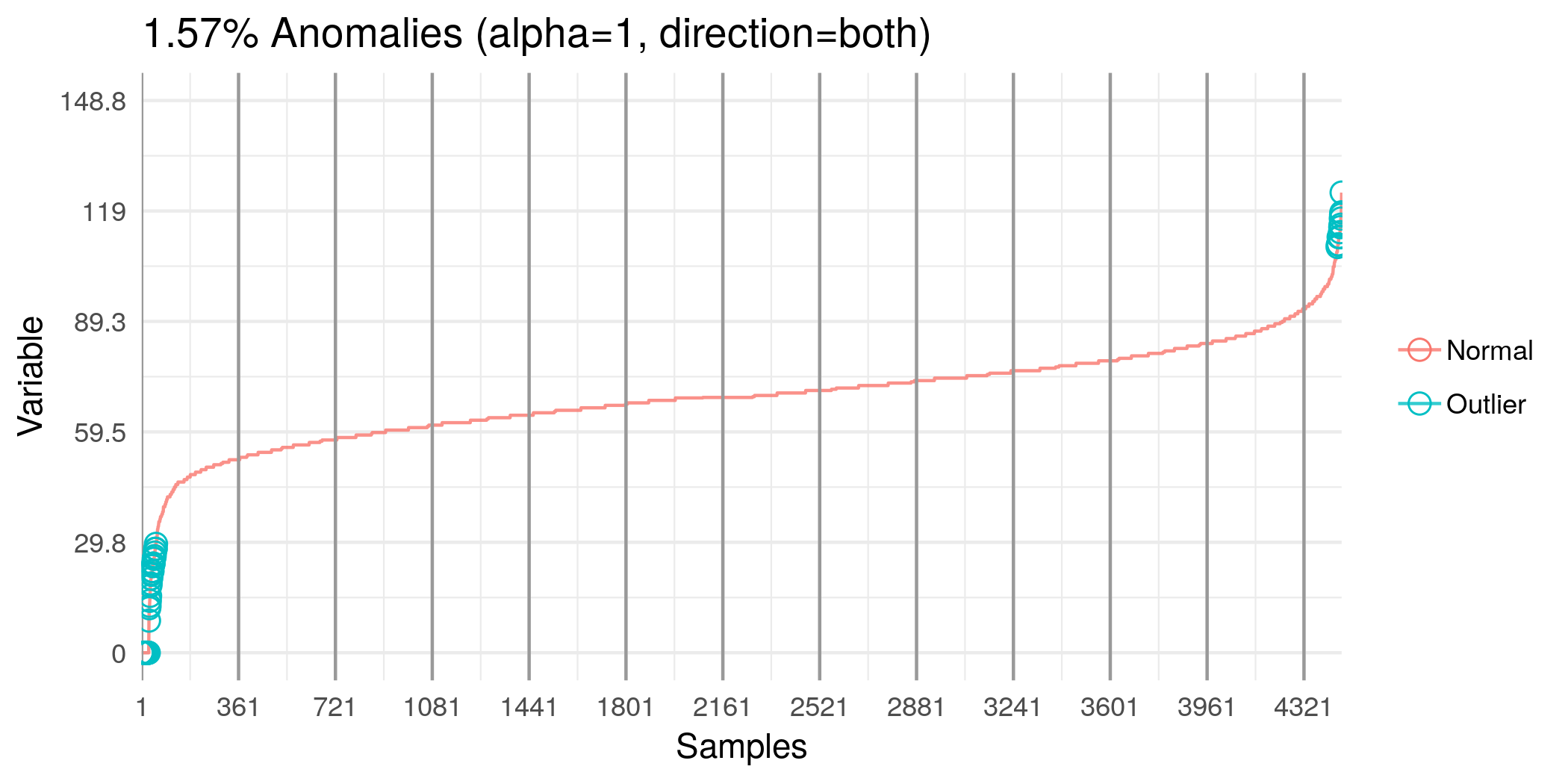}
    \caption{\textbf{Outlier Detection Algorithm}. After sorting data points in an increasing manner for the variable of interest, the outliers are detected after applying the anomaly detection algorithm (blue samples).}
    \label{out}
\end{figure}

The \code{data_preproc} function can be used to preprocess raw datasets. The function is designed to address outlier-detection and imputation of missing data. In order to find outliers, \code{data_preproc} uses an anomaly-detection algorithm from \citet{Vallis2014} for time series data (Fig. \ref{anom}). For each variable, it sorts the observations in a decreasing (or increasing) order and it defines the anomalies detected by the algorithm, as outlier data (Fig. \ref{out}). The function then removes the outliers and behave them as missing observations. 
The missing observations are imputed by the mean or the median of the whole set of observations for the measurement if it is continuous or categorical, respectively. The method gets a parameter,  \code{levels}, an integer value indicating the maximum number of levels of a categorical variable. The method returns the dataset with continuous variables as \code{numeric} and categorical variables as \code{factor} data types (see \ref{preprocess} for the practical example).


\subsection{Test Associations}
The function \code{test_pair} implements \textit{Pearson's Chi-squared}, \textit{ANOVA}, and \textit{correlation} tests for categorical-categorical, categorical-continuous, and continuous-continuous pairs of variables, respectively. One can easily use \code{test_pair} (\code{test_assoc}) in order to test any desired association between two (multiple) variables. Additionally, \code{test_assoc} implements  multiple hypothesis tests, using False Discovery Rate correction of Benjamini and Hochberg. 

\subsection{Plot Associations}
The function \code{plot_assoc} is implemented to facilitate single- and pairwise-variable visualizations. For a single continuous or categorical variable, it creates a bar plot and density plot, respectively. Pairwise-variable visualizations consist of a boxplot of the continuous variable for different levels of the categorical one, a scatter plot for two continuous variables and a heatmap illustrating the relation of different levels of two categorical variables. There is also a logical parameter \code{interactive} indicating if the output plot a \pkg{highcharter} object (\citet{highcharter}). It will output a print-friendly plot using \proglang{R} package \pkg{ggplot2} when the parameter is set to \code{False} (\citet{wickham2010ggplot2}).

\subsection{GMs}
As mentioned in \ref{graphical_models}, GMs are efficient for computation and interpretation of the whole structure of associations among the variables. However, each set of variables requires a proper GM. Package \pkg{muvis} implements three types of GMs: 
(i) \code{ggm} implements five different methods (i.e., AIC-based, BIC-based, partial correlation thresholding, significance testing, and glasso) for GGM structure learning, based on \proglang{R} packages \pkg{gRim}, \pkg{glasso}, \pkg{SIN}, and \pkg{gRbase}.
(ii) \code{dgm} is for constructing Directed (causal) GMs (DGMs) using \pkg{pcalg} package.
(iii) \code{min_forest} is designed for High-dimensional Graphical Modeling and is based on its implementation in the package \pkg{gRapHD} (\citet{Rosa2010}).
For all of the mentioned methods, there is a parameter \code{community} which indicates if the user wants to apply community detection algorithms on the graph to find the modules within the graph. To this end, we used Louvain method from \proglang{R} package \pkg{igraph} (\citet{Csardi2010,Blondel2008}).
There is another parameter \code{plot} which is a logical parameter indicating if the user wants to plot the graph. The \proglang{R} package \pkg{qgraph} is used to implement this graph visualization (\citet{Epskamp2012}).

\subsection{KL-based Functions}
 The functions \code{VKL} and \code{VVKL} implement the KL-methods. The parameter \code{permute} indicates the number of permutations as described in \ref{kl_pval} and is used to find the significance of the KL values. We used the function \code{KL.plugin} from R package \code{entropy} to calculate KL values (\citet{Hausser2014}). As this function works on discrete data, prior to calculating KL for continuous variables, we discretize and consider them like discrete data (see \ref{vkl}).

\section{The NHANES 2005-2006 dataset}
\label{sec_data}
The National Health and Nutrition Examination Surveys (NHANES) (\citet{nhanes}) is a program of studies about health and nutrition for US residents. We examined the functionality of \pkg{muvis} on NHANES 2005-2006 dataset which contains 7449 variables and 10,348 samples (see \url{https://www.icpsr.umich.edu/icpsrweb/ICPSR/studies/25504/summary} for more details).\\
Based on the number of missing values, we selected 161 variables including one ID, 74 continuous, and 86 categorical variables (having two to fifteen levels) and 4461 individuals (samples) aged from 20 to 85 years, including about 1\% missing values. See Table \ref{vars} for description of the variables that are mentioned in this article. 
Complete list of 161 variables is available in \code{NHANES} dataset of the package. The next parts describe the analysis of this dataset using \pkg{muvis}.

\subsection{Loading the package and the data}
\pkg{muvis} is available at \url{https://github.com/bAIo-lab/muvis}. Once the package is installed and loaded into the \proglang{R} environment, the \code{NHANES} dataset can be loaded as a \code{dataframe} by a call to \code{data} as in the following.
\begin{Schunk}
\begin{Sinput}
> library(muvis)
> data("NHANES")
\end{Sinput}
\end{Schunk}
\subsection{Preprocessing the data} \label{preprocess}
\begin{figure}
    \begin{subfigure}[b]{0.5\textwidth}
    \includegraphics[width=\textwidth]{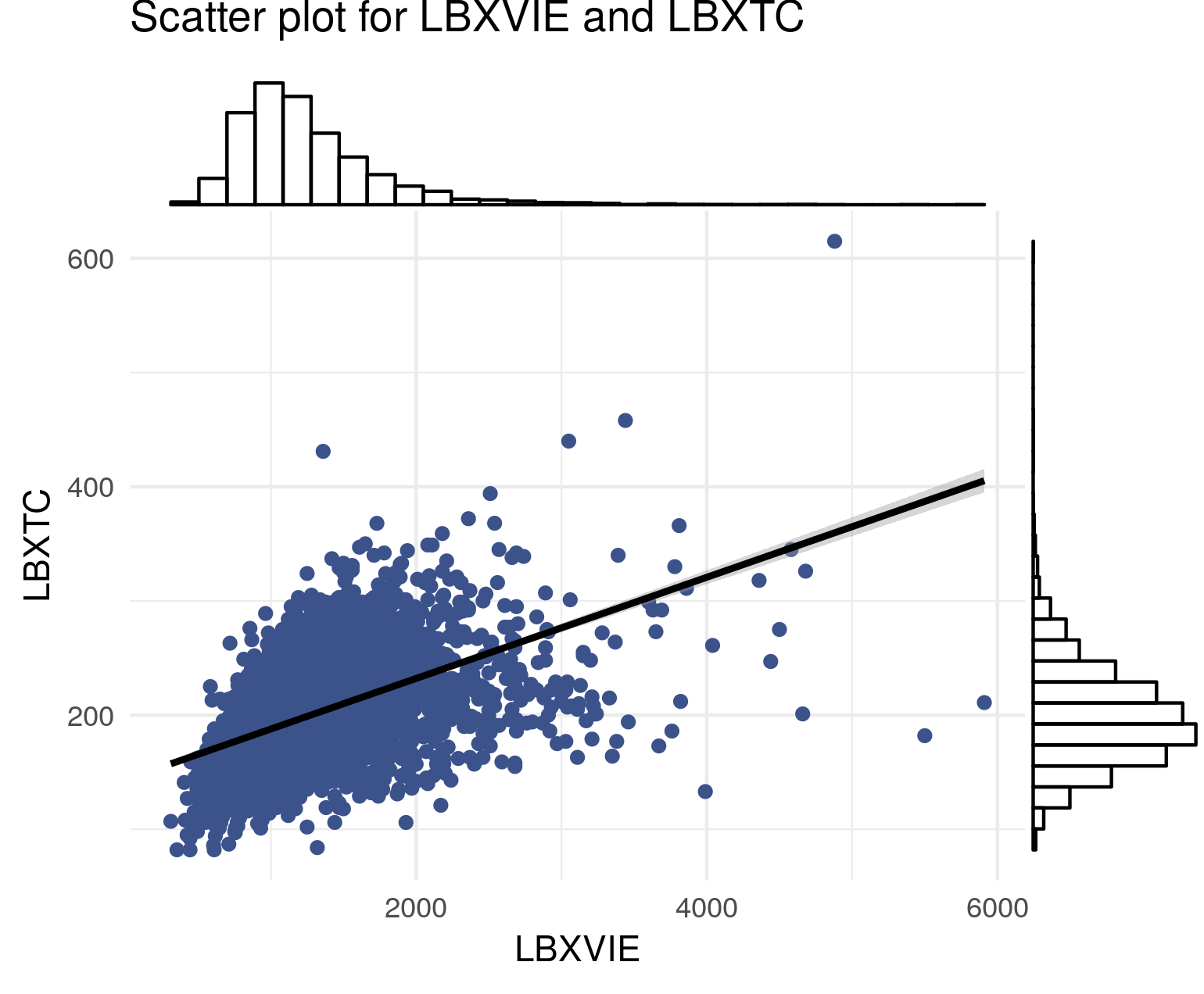} 
    \caption{}
    \label{fig11:subim1}
    \end{subfigure}
    \begin{subfigure}[b]{0.5\textwidth}
    \includegraphics[width=\textwidth]{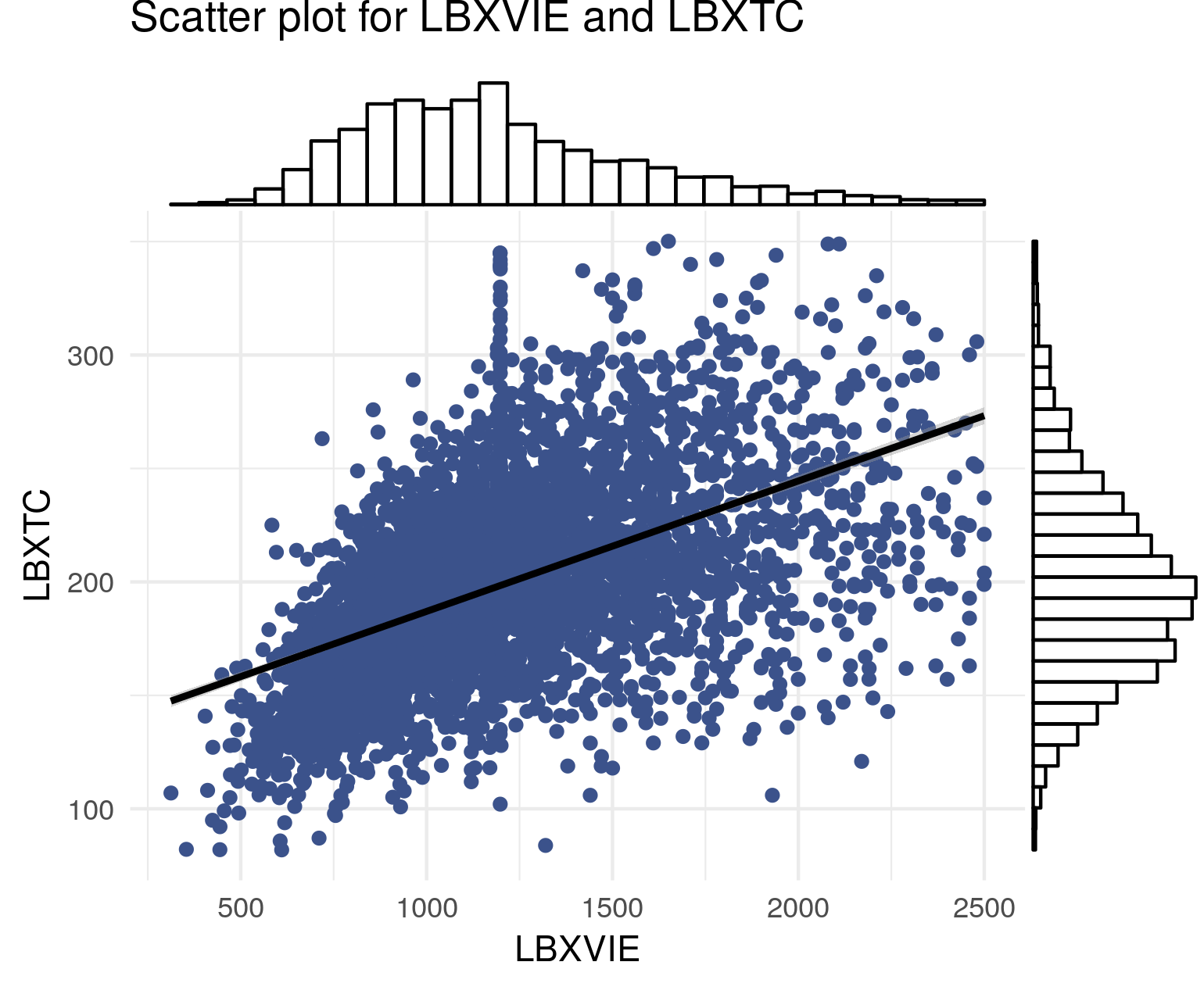} 
    \caption{}
    \label{fig11:subim2}
    \end{subfigure}
  \caption{\textbf{Scatter plot} of \code{LBXTC} (total cholesterol. mg/dL) and \code{LBXVIE} (vitamin E. ug/dL) for NHANES dataset (a) with outliers and (b) without outliers. Each blue point indicates a sample and the black line shows the regression line fitted on two variables.}
  \label{fig:sfigoutlier}
\end{figure}

We use \code{data_preproc} function for preprocessing of \code{NHANES}. 
The \code{detect.outliers} option is used to exclude outliers for each variable. As a first step through interpretation, we plot the relation between \code{LBXVIE} (a variable in the dataset indicating the amount of vitamin E) and \code{LBXTC} (the amount of total cholesterol) with and without outliers. The difference has been shown in Fig. \ref{fig:sfigoutlier}.
\begin{Schunk}
\begin{Sinput}
> nhanes_with_outliers <- data_preproc(NHANES, levels = 15, alpha = 0.5)
> nhanes <- data_preproc(NHANES, levels = 15, detect.outliers = TRUE, 
alpha = 0.5)
> plot_assoc(nhanes_with_outliers, vars = c("LBXVIE", "LBXTC"))
> plot_assoc(nhanes, vars = c("LBXVIE", "LBXTC"))
\end{Sinput}
\end{Schunk}

\subsection{GGM for continuous data} \label{ggm}
\begin{figure}
    \centering
    \includegraphics[scale=.7]{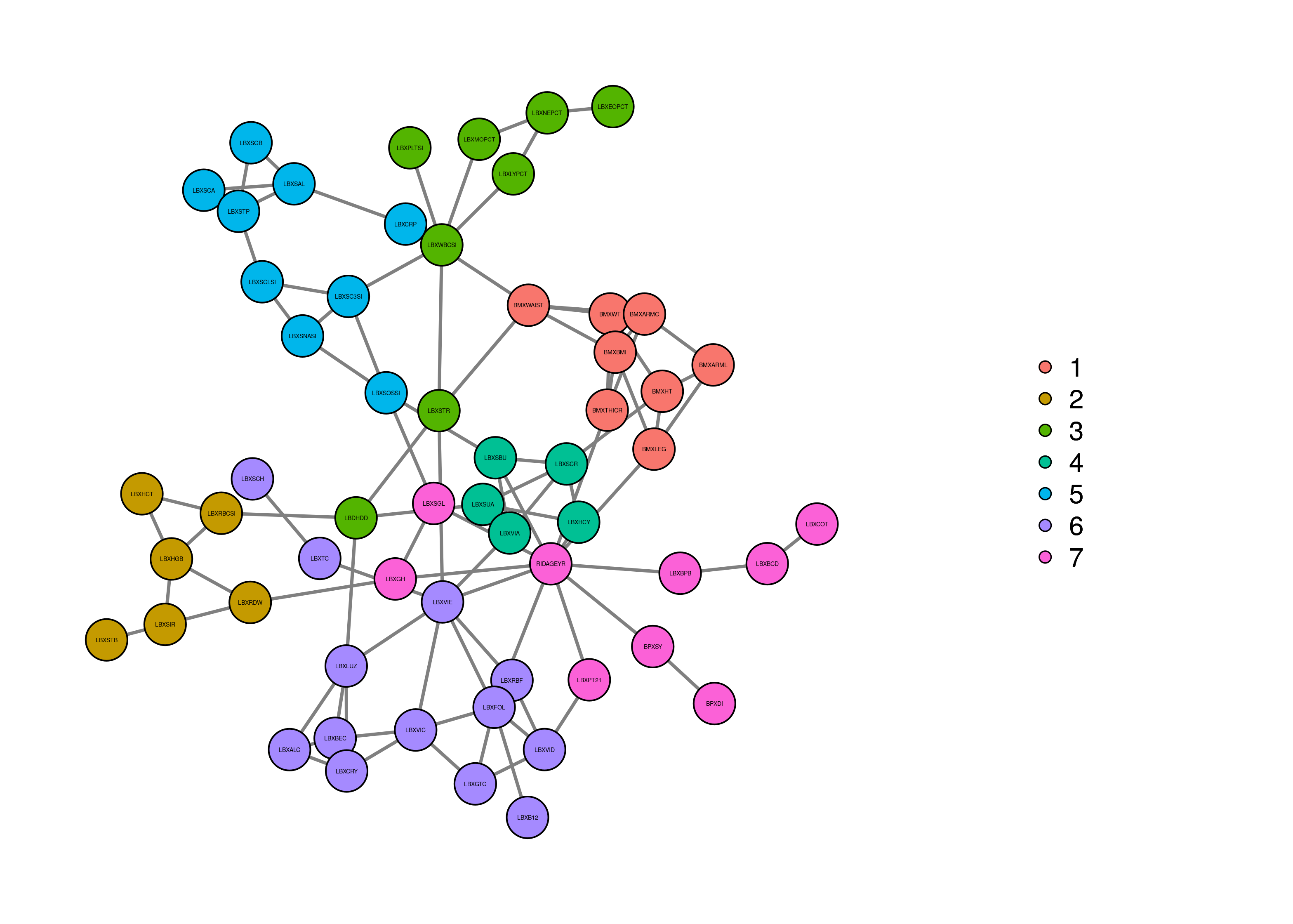}
    \caption{\textbf{GGM}. The graph is constructed using \code{ggm} function with the intersection of the models estimated by \code{sin} and \code{glasso} methods. The largest connected component of the graph is shown. The nodes represent continuous variables in the data and are colored according to their community.}
    \label{fig:ggm}
\end{figure}

We construct a GGM for continuous variables using \code{ggm} function. In this example, we construct it by intersecting \code{glasso} and \code{sin} algorithms. The largest connected component of the estimated graph is visualized by \code{graph_vis} function (Fig. \ref{fig:ggm}). 
\begin{Schunk}
\begin{Sinput}
> nhanes$SEQN <- NULL
> nhanes_ggm <- ggm(nhanes, significance = 0.05,
rho = 0.15, community = TRUE, methods = c("glasso", "sin"), plot = F)
> grph_clustrs <- clusters(nhanes_ggm$graph)
> new_ggm <- induced.subgraph(nhanes_ggm$graph,
V(nhanes_ggm$graph)[which(grph_clustrs$membership == which.max(grph_clustrs$csize))])
> ggm_vis <- graph_vis(new_ggm, plot = T, 
filetype = "png", filename = "nhanes_ggm")
\end{Sinput}
\end{Schunk}

Investigating each community one can appraise the efficiency of the model: Community 1 are all about body measurements (e.g., BMI, waist circumstance, height, etc.); community 2 explains red blood cell profile; community 3 contains measurements about different types of white blood cells; community 4 accommodates uric acid, vitamin A, Urea, Creatinine, and Homocysteine; community 5 includes variables describing body biochemistry profile; community 6 contains some vitamins and chemical compounds (e.g. carotenoids, folate, etc.); and finally community 7 contains age, blood pressure, lead, and Parathyroid hormone.

\subsection{Causal network for continuous data} \label{dgm}
\begin{figure}
    \centering
    \includegraphics{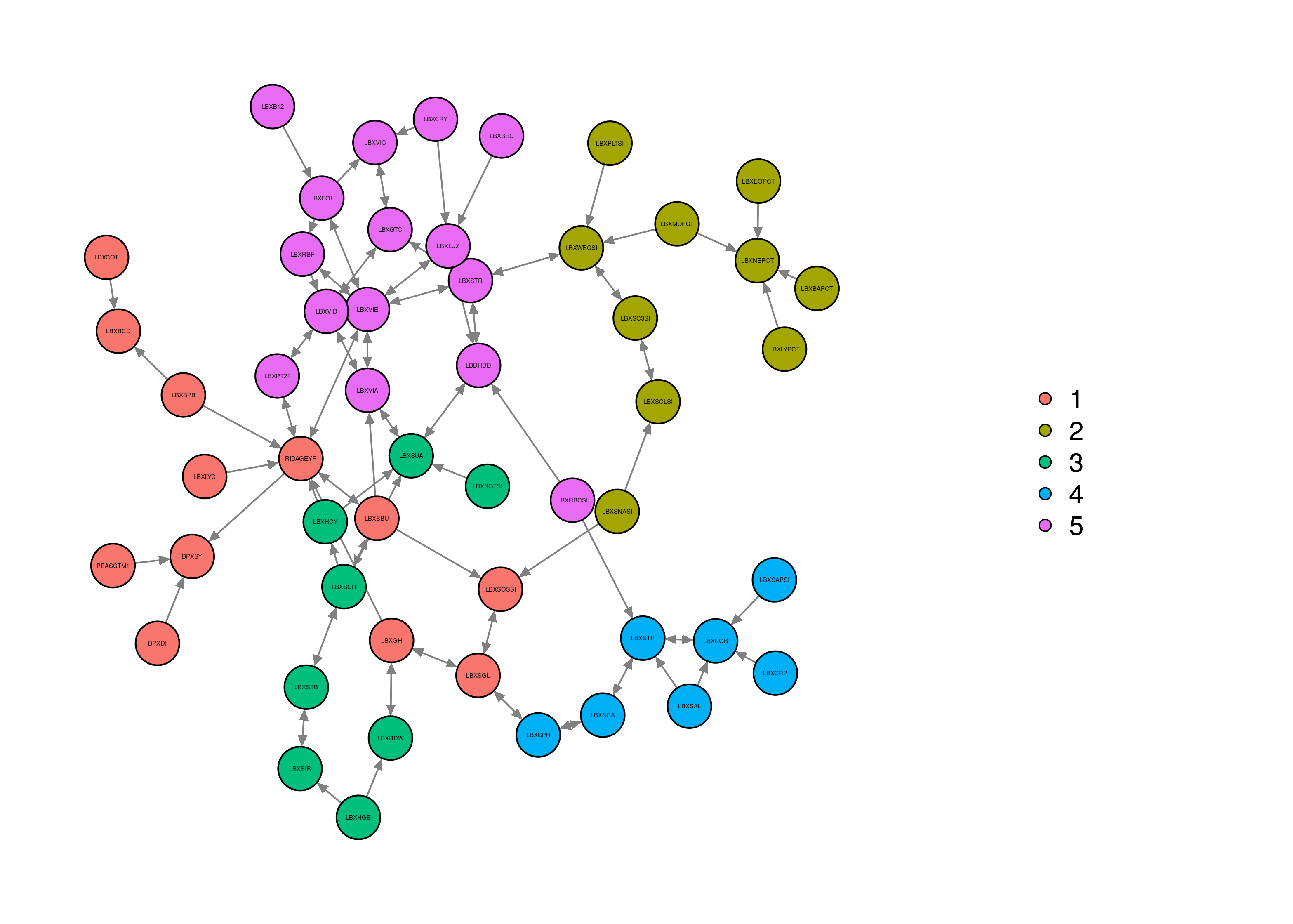}
    \caption{\textbf{Causal graph}. The largest connected component of the graph is shown. Nodes indicate variables in the dataset and are colored based on their community number. Edges are directed based on the estimated causal relationship so that each edge is directed from cause to effect.}
    \label{fig:dgm}
\end{figure}

The causal (directed) network of continuous variables is constructed using \code{dgm} function with parameter \code{dtype = "gaussian"}. The largest connected component of the estimated graph is shown in Fig. \ref{fig:dgm}.
\begin{Schunk}
\begin{Sinput}
> nhanes_dgm <- dgm(nhanes, dtype = "gaussian", alpha = 1e-15)
> grph_clustrs <- clusters(nhanes_dgm$graph)
> new_dgm <- induced.subgraph(nhanes_dgm$graph,
V(nhanes_dgm$graph)[which(grph_clustrs$membership == which.max(grph_clustrs$csize))])
> dgm_vis <- graph_vis(new_dgm, plot = T, directed = T, filename = "nhanes_dgm",
filetype = "png")
\end{Sinput}
\end{Schunk}

\subsection{Minimal forest for mixed data} \label{mf}
\begin{figure}
\begin{subfigure}{\textwidth}
\centering
\includegraphics[width=0.7\textwidth]{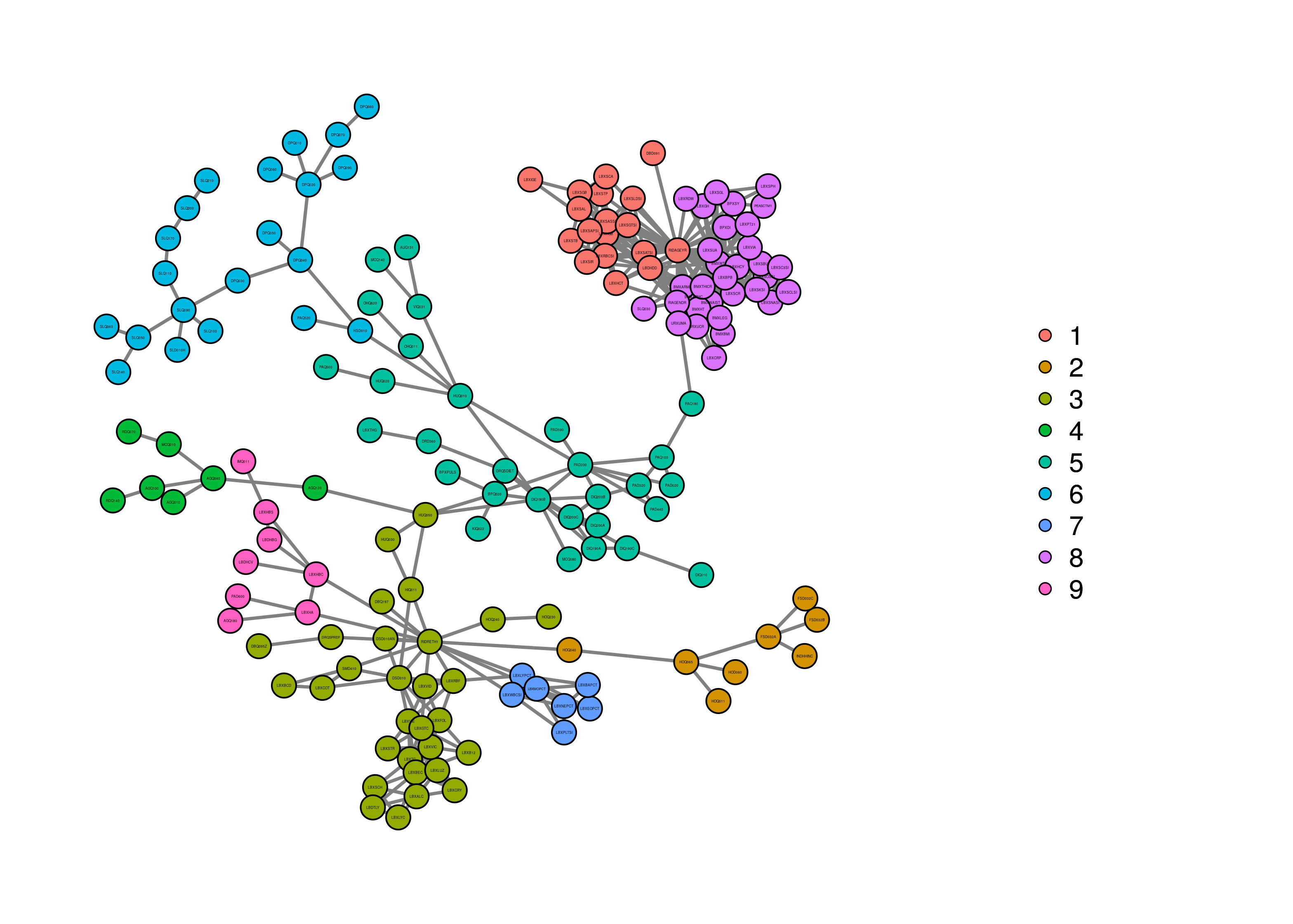}
\caption{}
\label{fig:subim8}
\end{subfigure}

\begin{subfigure}[b]{0.3\textwidth}
\includegraphics[width=\textwidth]{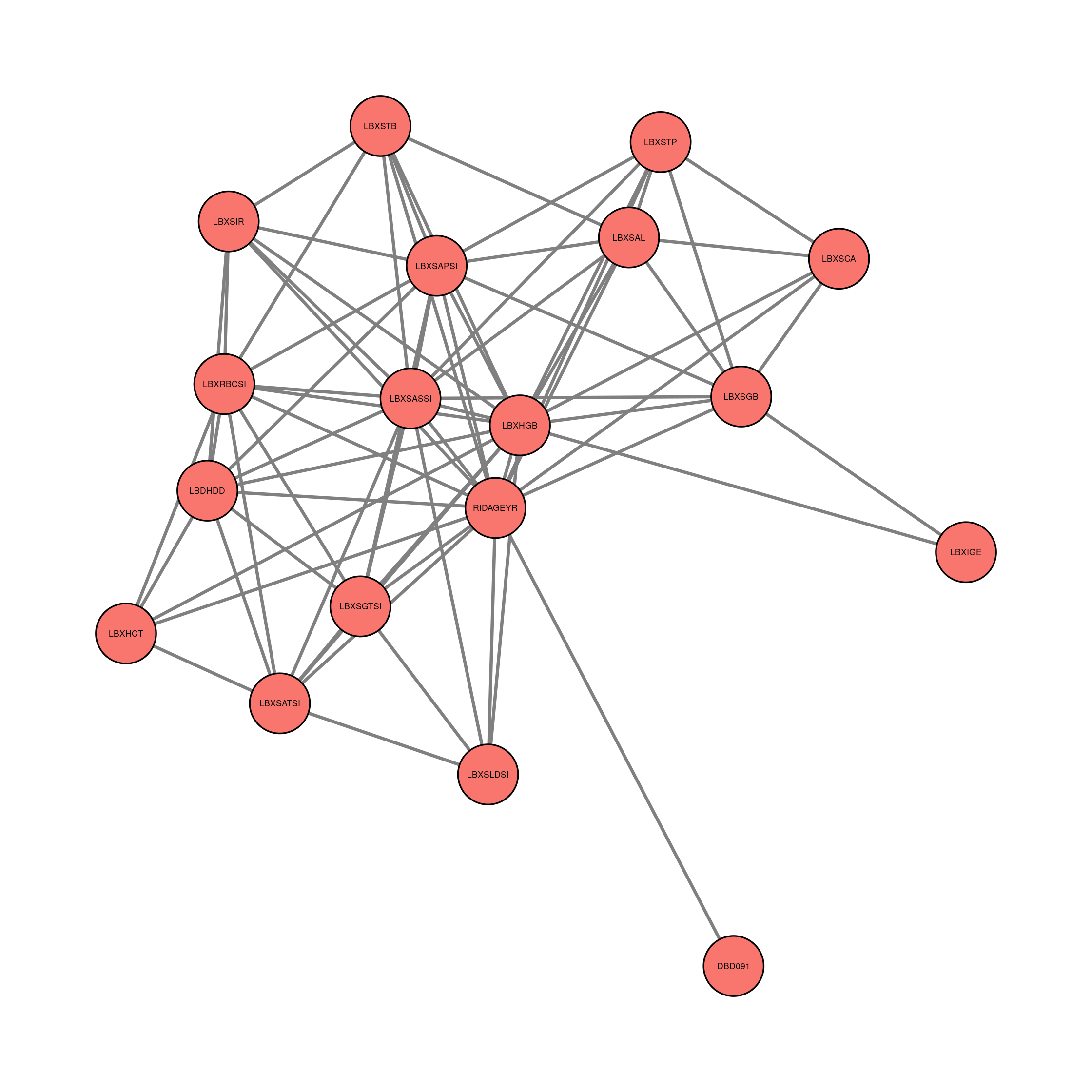} 
\caption{}
\label{fig:subim1}
\end{subfigure}
~
\begin{subfigure}[b]{0.3\textwidth}
\includegraphics[width=\textwidth]{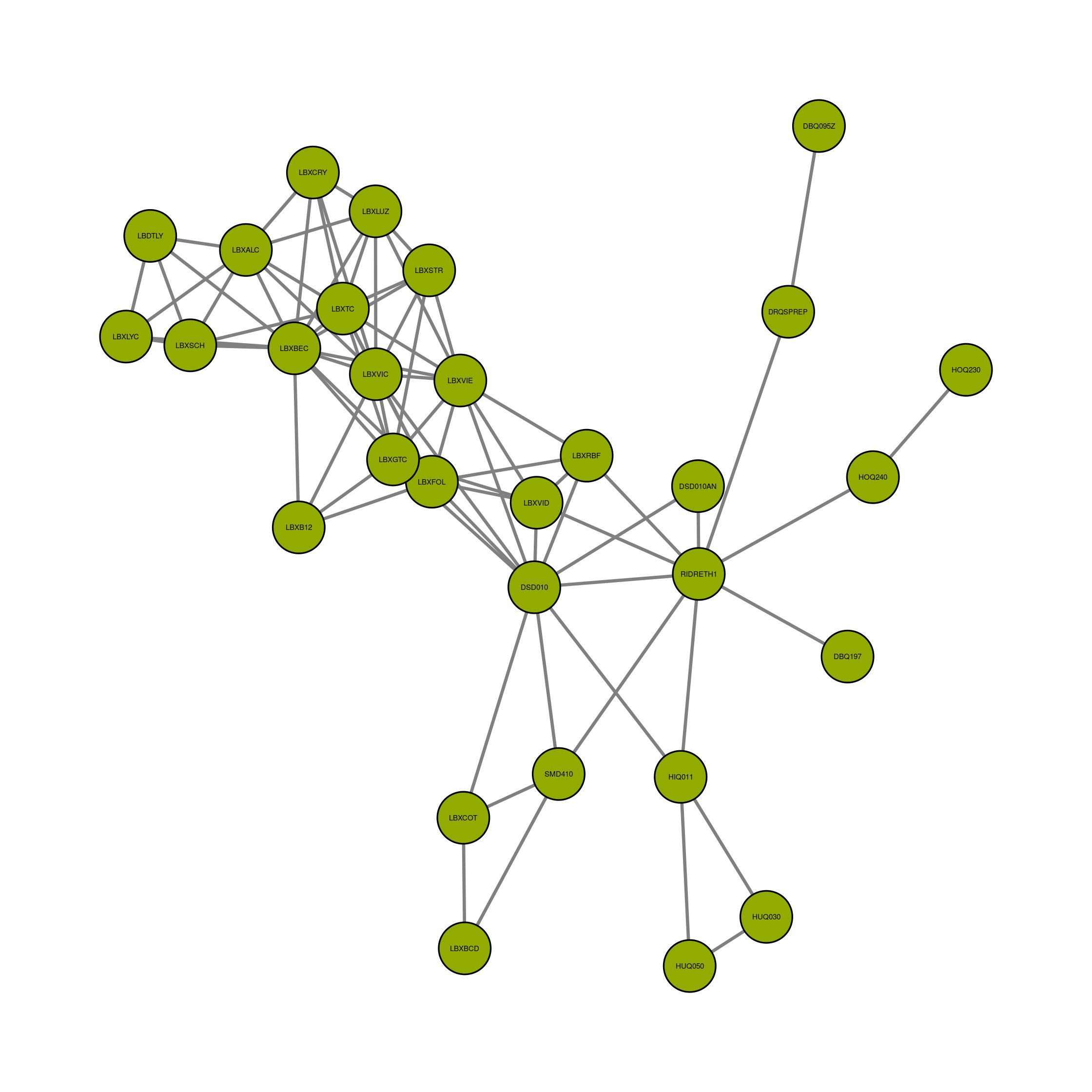} 
\caption{}
\label{fig:subim3}
\end{subfigure}
~
\begin{subfigure}[b]{0.3\textwidth}
\includegraphics[width=\textwidth]{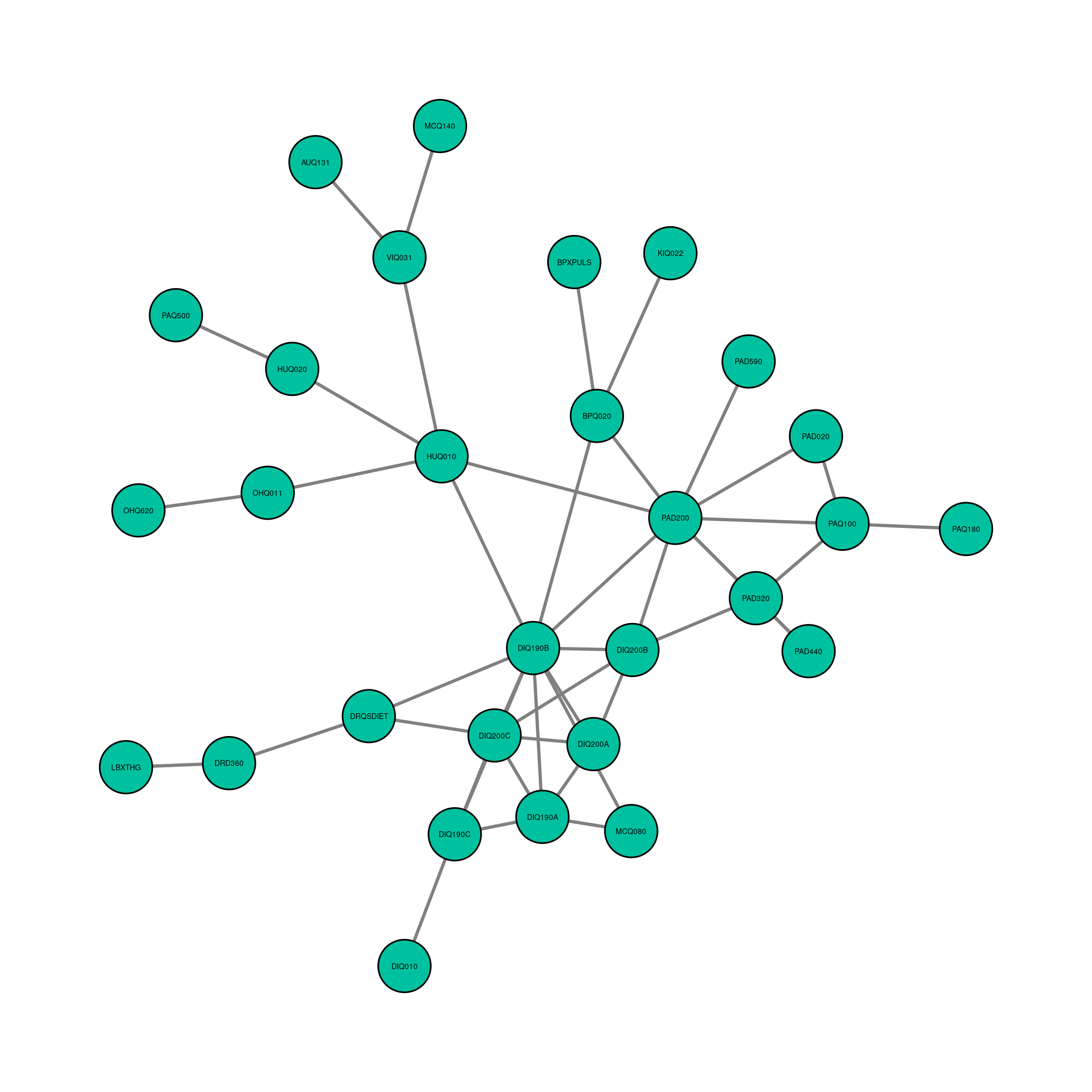}
\caption{}
\label{fig:subim5}
\end{subfigure}

\begin{subfigure}[b]{0.3\textwidth}
\includegraphics[width=\textwidth]{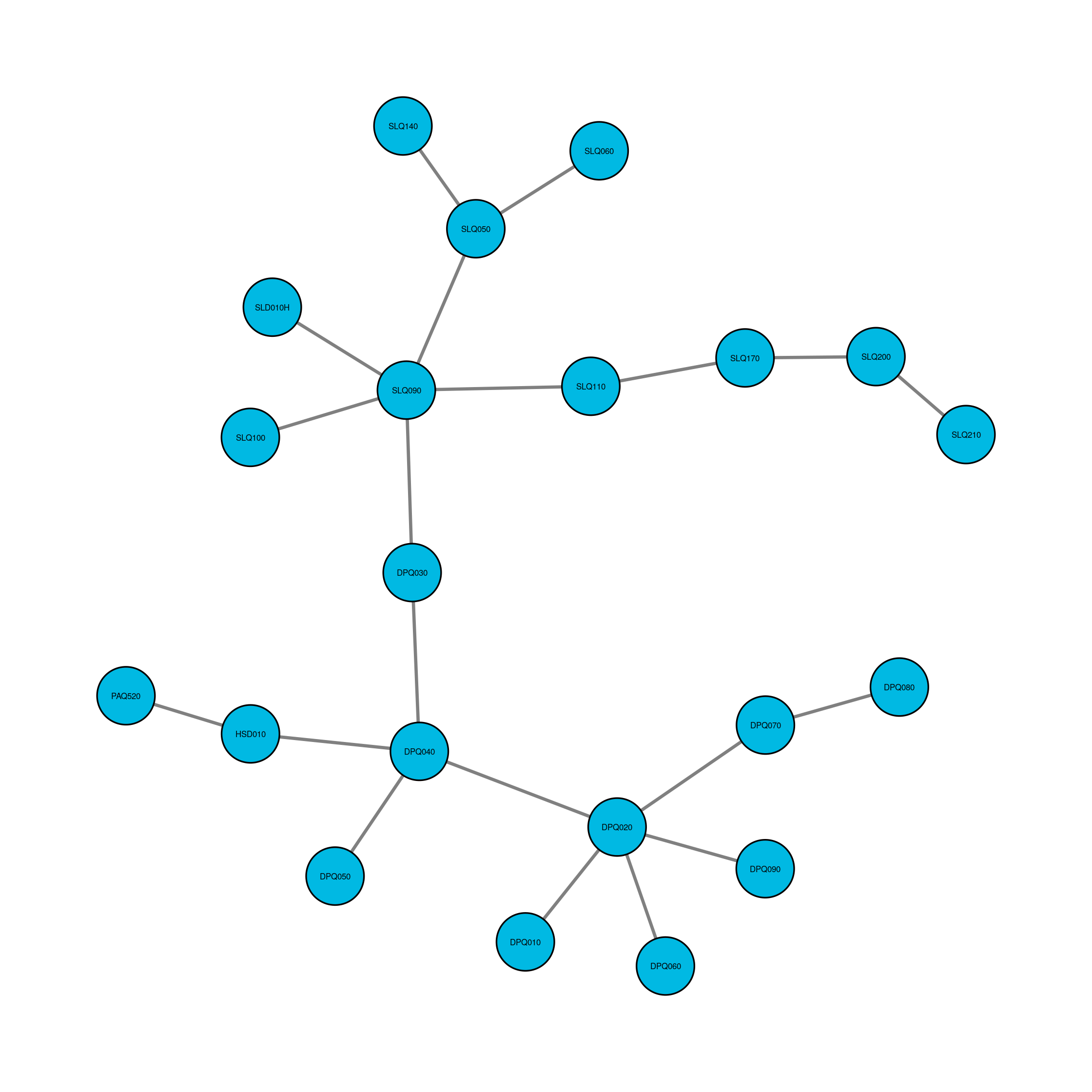}
\caption{}
\label{fig:subim6}
\end{subfigure}
~
\begin{subfigure}[b]{0.3\textwidth}
\includegraphics[width=\textwidth]{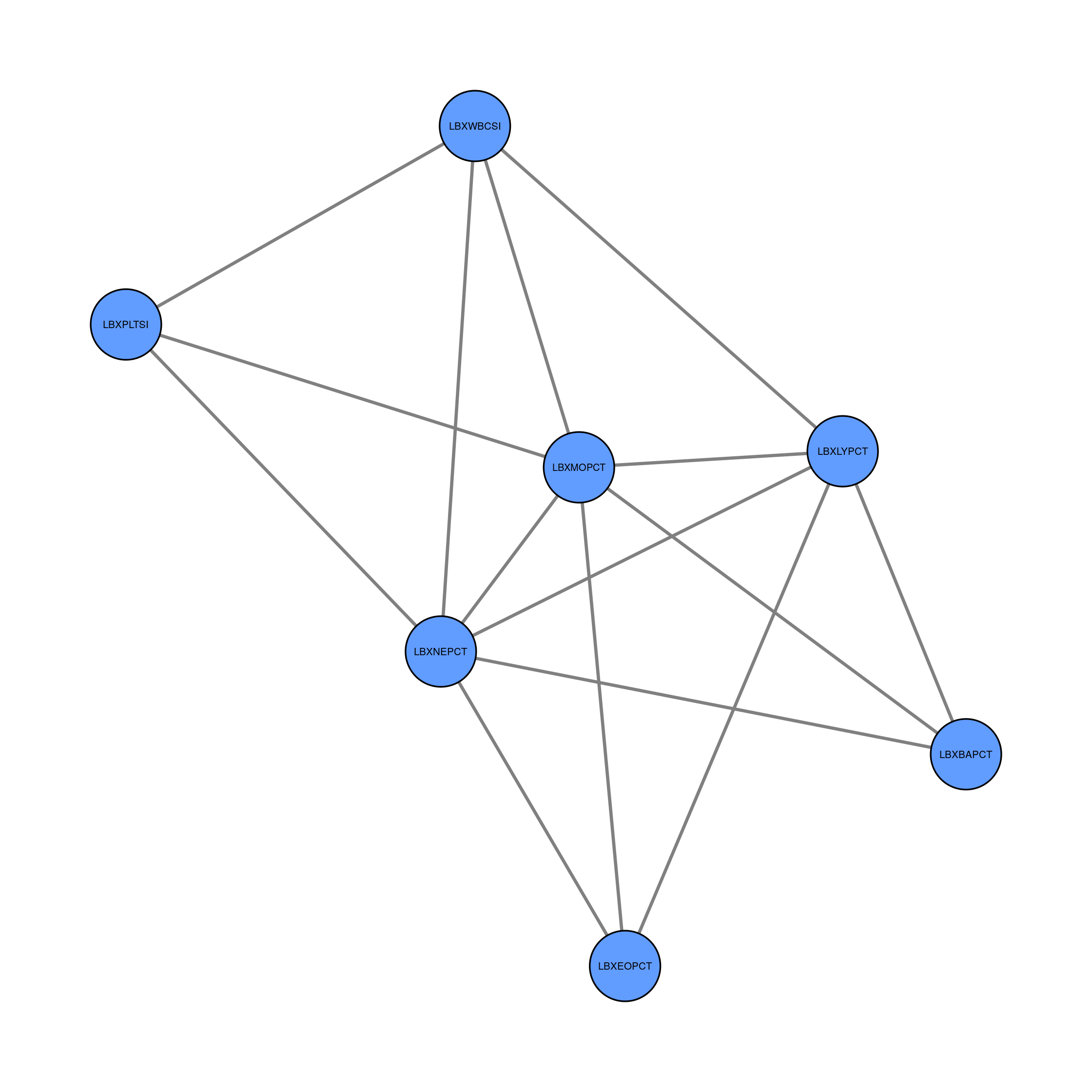}
\caption{}
\label{fig:subim7}
\end{subfigure}
~
\begin{subfigure}[b]{0.3\textwidth}
\includegraphics[width=\textwidth]{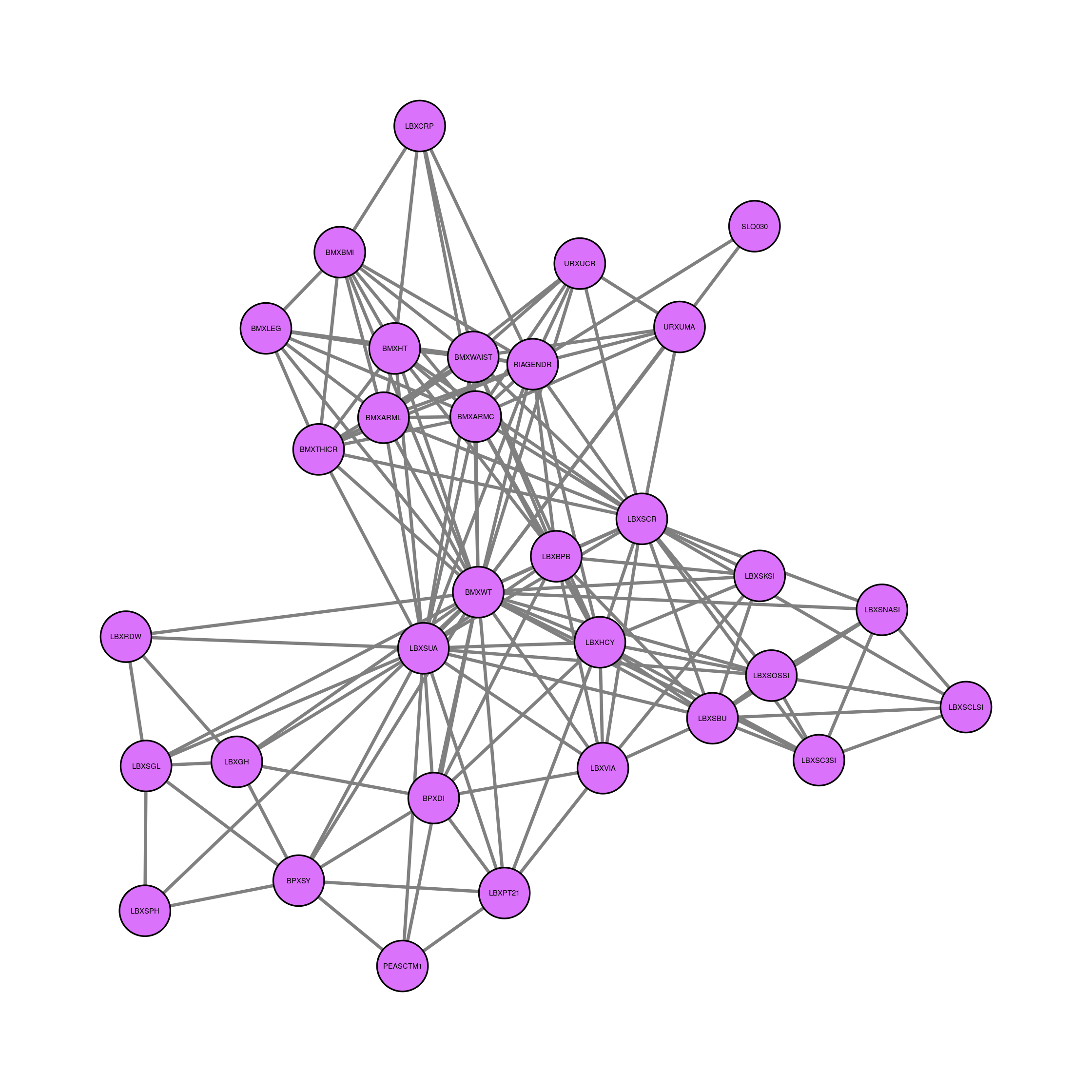}
\caption{}
\label{fig:subim8}
\end{subfigure}

\caption{\textbf{Minimal forest}. (a) The graph is constructed using \code{min\_forest} function with \code{BIC} method. The nodes indicate variables in the dataset and are colored with respect to their community. (b) community 1: red blood and body biochemistry profile. (c) community 3: chemical compounds and vitamins. (d) community 5: physical activity, health condition, and diabetes. (e) community 6: depression and sleeping status. (f) community 7: white blood cells. (g) community 8: body measurements and blood pressure.}
\label{fig:image2}  
\end{figure}

Using \code{min_forest} function we estimate the minimal forest with \code{BIC} method and detect communities in the graph. 
The estimated minimal forest and some of its communities are illustrated in Fig. \ref{fig:image2}.
\begin{Schunk}
\begin{Sinput} 
> nhanes_mf <- min_forest(nhanes, stat = "BIC", community = T, plot = F)
\end{Sinput}
\end{Schunk}



\begin{Schunk}
\begin{Sinput}
> mf_vis <- graph_vis(nhanes_mf$graph, plot = T, filetype = "png", 
filename =  "nhanes_mf_bic", plot.community = T)
\end{Sinput}
\end{Schunk}

\subsection{Community analysis of the minimal forest}
In the following section, we inspect communities 1, 3, 5, 6, 7, and 8 of the minimal forest graph to demonstrate the functionally of the algorithm.

\begin{figure}
    \begin{subfigure}[t]{0.5\textwidth}
    \includegraphics[width=\textwidth]{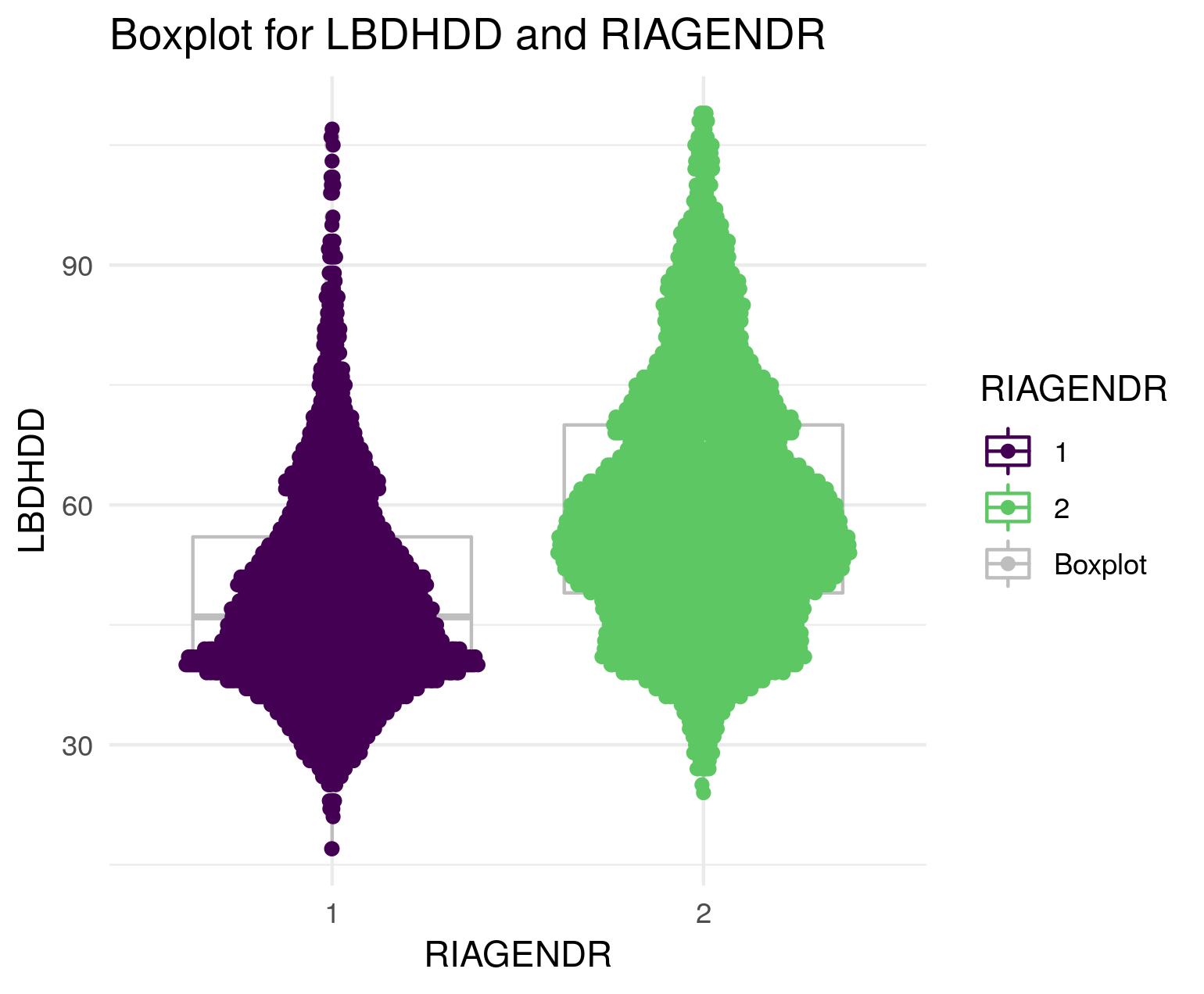}
    \caption{}
    \label{fig:hddgend}
    \end{subfigure}
    ~
    \begin{subfigure}[t]{0.5\textwidth}
      \includegraphics[width=\textwidth]{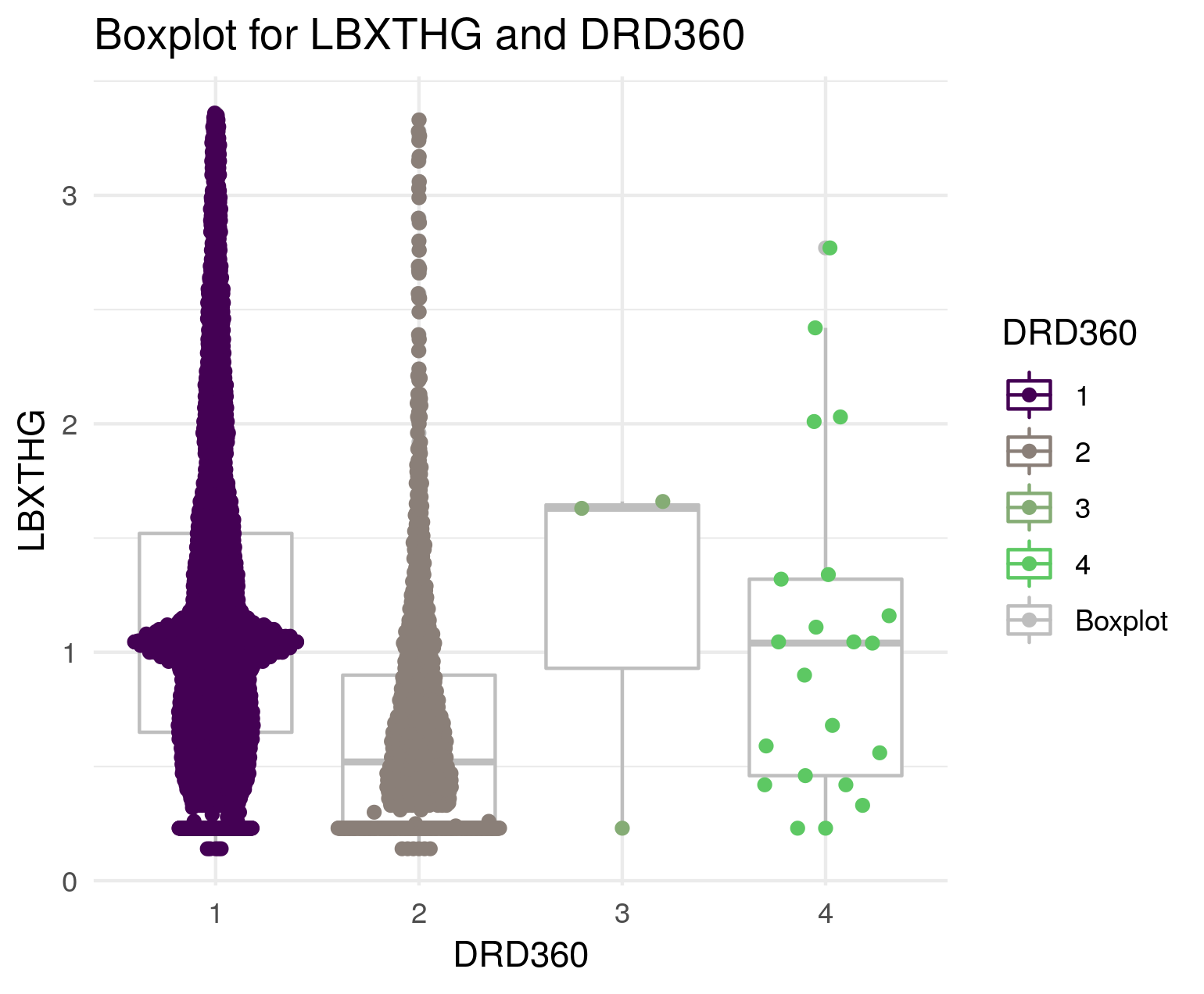}
      \caption{}
      \label{fig:fish}
    \end{subfigure}\\[1ex]
    ~
    \begin{subfigure}[t]{\textwidth}
    \centering
      \includegraphics[width=.5\textwidth]{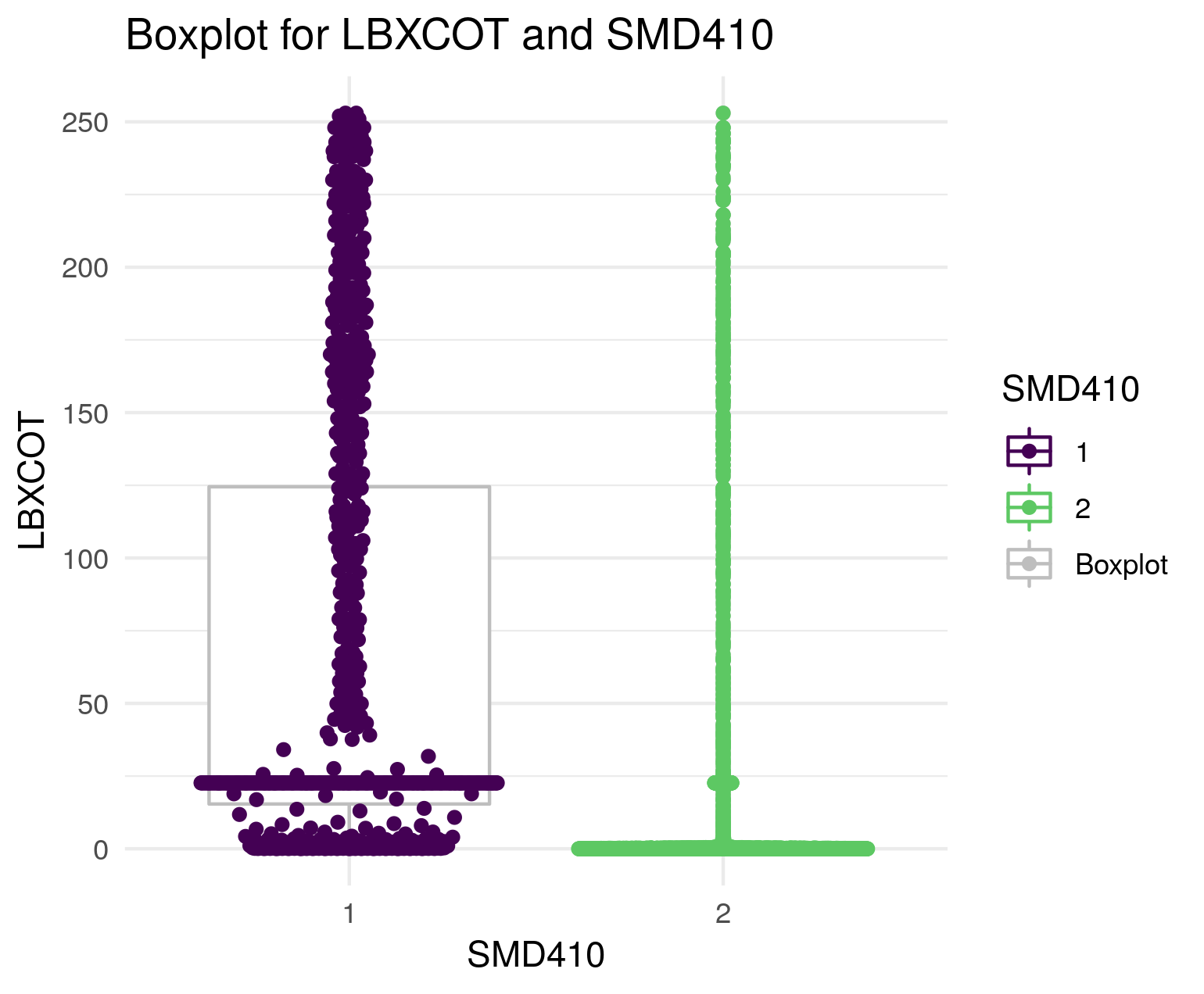}
      \caption{}
      \label{fig:smd_cot}
    \end{subfigure}
    \caption{\textbf{Boxplot} of (a) \code{LBDHDD} (direct HDL-cholesterol. mg/dL) and \code{RIAGENDR} (gender. 1 for Male and 2 for Female), (b) \code{LBXTHG} (total mercury in the blood. ug/L) and \code{DRD360} (eating fish in past 30 days. 1 for Yes, 2 for No, 3 for Refused and 4 for Don't know), (c) \code{LBXCOT} (cotinine. ng/mL) and \code{SMD410} (smoking in household. 1 for Yes and 2 for No). The plots are colored according to the different levels of their categorical variable. points in these plots represent samples in the data.}
    \label{plots}
\end{figure}

Most of the nodes in community 1 (Fig. \ref{fig:subim1}) are related to red blood cells and body biochemistry profile. There is an association between \code{LBDHDD} (direct HDL-cholesterol) and \code{RIAGENDR} (gender) in this community, shown in Fig. \ref{fig:hddgend}.

\begin{CodeChunk}
\begin{CodeInput}
> signif <- nhanes_mf$significance
> signif[signif$edges.from == 'RIAGENDR' & signif$edges.to == 'LBDHDD', ]

\end{CodeInput}
\begin{CodeOutput}
   edges.from edges.to statistics      p.value
   RIAGENDR   LBDHDD   619.5036 1.737264e-13
\end{CodeOutput}
\begin{CodeInput}
> plot_assoc(nhanes, vars = c("LBDHDD", "RIAGENDR"))
\end{CodeInput}
\end{CodeChunk}

\begin{figure}
    \begin{subfigure}[b]{0.5\textwidth}
      \includegraphics[width=\textwidth]{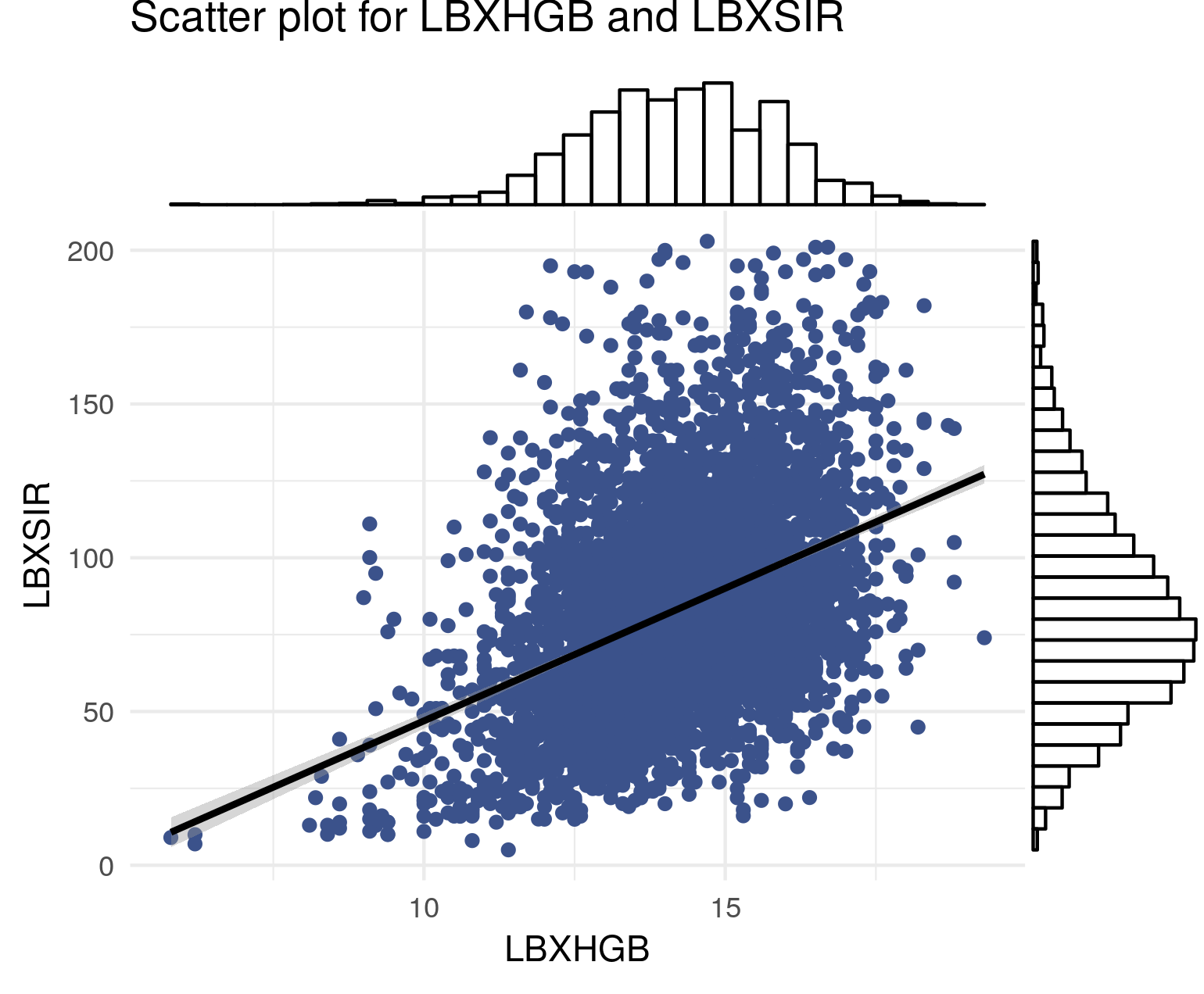}
      \caption{}
      \label{fig:hgb_sir}
    \end{subfigure}
    ~
    \begin{subfigure}[b]{0.5\textwidth}
      \includegraphics[width=\textwidth]{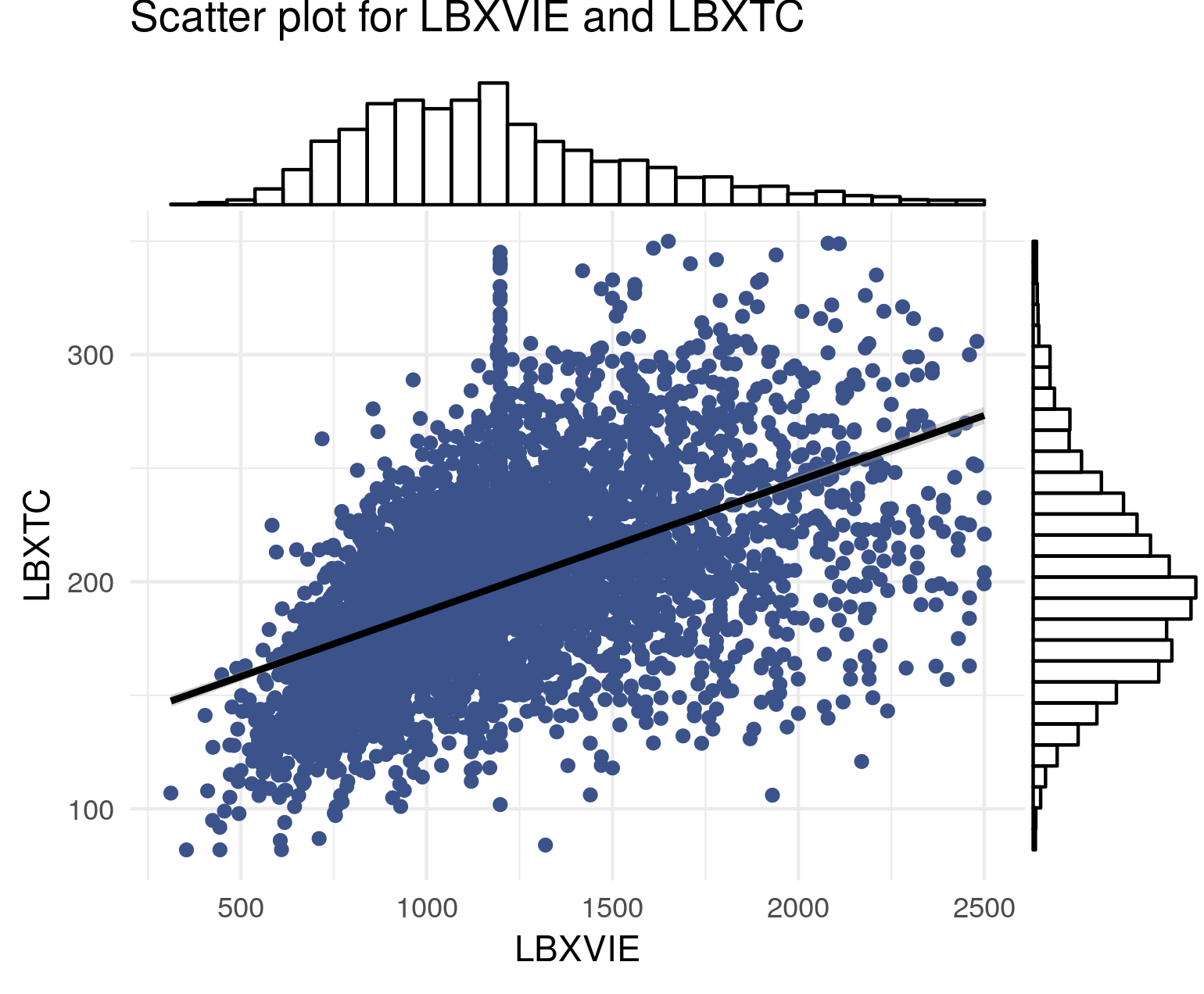}
      \caption{}
      \label{fig:tc_vie}
    \end{subfigure}
        \caption{\textbf{Scatter plot} of (a) \code{LBXHGB} (hemoglobin. g/dL) and \code{LBXSIR} (refrigerated iron in blood. ug/dL), (b) \code{LBXTC} (total cholesterol. mg/dL) and \code{LBXVIE} (vitamin E. ug/dL). Each point in the plots indicates a sample in the data.}
    \label{fig:my_label}
\end{figure}

Fig. \ref{fig:hgb_sir}  shows another relation in community 1, which is between nodes \code{LBXHGB} (hemoglobin) and \code{LBXSIR} (refrigerated iron in blood).
Community 3 (Fig. \ref{fig:subim3}) mostly includes chemical compounds and vitamins (i.e. E, C, B12, and D). As an example, the scatter plot of two variables \code{LBXVIE} (vitamin E) and \code{LBXTC} (totoal cholesterol) is illustrated in Fig. \ref{fig:tc_vie}. There are also two connected nodes \code{SMD410} (smoking in household) and \code{LBXCOT} (cotinine) in this community that their relation is shown in Fig. \ref{fig:smd_cot}.
Nodes in community 5 (Fig. \ref{fig:subim5}) indicate three groups of variables, namely, physical activity, health condition, and diabetes. There is a chain relation that connects \code{DRD360} (eating fish in past 30 days) to diabetes group through node \code{DRQSDIET} (being on a special diet). \code{DRD360} is also connected to \code{LBXTHG} (total mercury in blood). The relation is plotted in Fig. \ref{fig:fish}.
\code{DPQ030} (trouble with sleeping) connects sleep disorder nodes to variables about depression in community 6 (Fig. \ref{fig:subim6}).
Community 7 (Fig. \ref{fig:subim7}) consists of different types of white blood cells. Body measurement variables and blood pressure are located in community 8 (Fig. \ref{fig:subim8}).

\subsection{Variable-wise KL-divergence} \label{vkl}
Focusing on the variable \code{PAD590} (TV usage) we select two groups of samples: (i) The participants who watch TV less than an hour (g1) and (ii) those who watch more than 5 hours (g2) a day . We use \code{VKL} function with \code{permute = 1000} so as to find the most different features between these two groups. In the following code, we get five variables with the highest KL-divergence values excluding \code{PAD590}.
\begin{CodeChunk}
\begin{CodeInput}
> g1 <- which(nhanes$PAD590 == 1)
> g2 <- which(nhanes$PAD590 == 6)
> KL <- VKL(nhanes, group1 = g1, group2 = g2, permute = 1000)
> KL[2:6, ]
\end{CodeInput}
\end{CodeChunk}
\begin{CodeChunk}
\begin{CodeOutput}
                KL variable p.value
HSD010   0.2581593   HSD010   0.001
PAQ520   0.2193065   PAQ520   0.001
HUQ010   0.2157753   HUQ010   0.001
RIDAGEYR 0.2094270 RIDAGEYR   0.001
LBXALC   0.2066657   LBXALC   0.001
\end{CodeOutput}
\end{CodeChunk}
\code{HSD010} (general health condition) is the most different variable between \code{g1} and \code{g2} based on KL-divergence values. See Table \ref{vars} for description of the other variables.

\subsection{Violating Variable-wise Kulback-Leibler Divergence}
\begin{figure}
    \centering
         \includegraphics[width=0.5\textwidth]{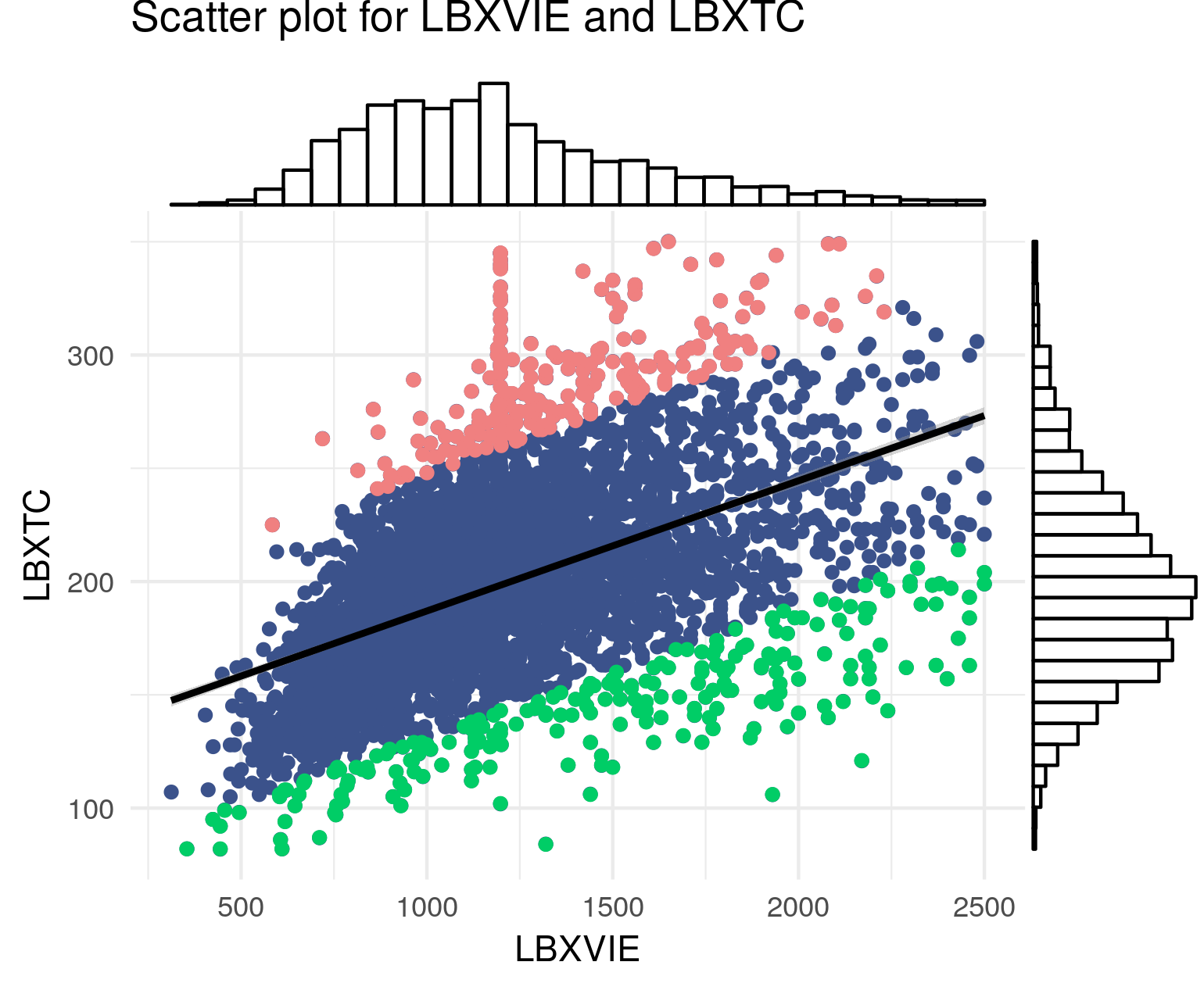}
        \caption{\textbf{Scatter plot} of \code{LBXTC} (total cholesterol. mg/dL) and \code{LBXVIE} (vitamin E. ug/dL). A line is fitted to the variables and the samples with absolute residual values within the highest 5 percentile are considered as the outliers. The two outlier groups are colored with red and green colors. }
        \label{fig:kl_outlier}
\end{figure}

 As Fig. \ref{fig:tc_vie} shows, there is an approximately linear relationship between vitamin E and total cholesterol. Using \code{VVKL} function with \code{permute = 100}, we find the most important (categorical) variables violating such linear relationship. The result suggests \code{DSD010} (taking dietary supplements) and \code{DRQSPREP} (type of table salt used) as variables with the highest KL-divergence. Two groups of samples with the highest absolute residual values (top 5 \%) with respect ot the fitted line are remarked as outliers and highlighted in Fig. \ref{fig:kl_outlier}.
\begin{CodeChunk}
\begin{CodeInput}
>  KL <- VVKL(nhanes[, 75:160], var1 = nhanes$LBXVIE, var2 = nhanes$LBXTC, 
plot = T, var1.name = "LBXVIE", var2.name = "LBXTC", permute = 100)
> head(KL$kl)
> KL$plot
\end{CodeInput}
\end{CodeChunk}
\begin{CodeChunk}
\begin{CodeOutput}
                KL variable p.value
DSD010   0.1485264   DSD010    0.01
DRQSPREP 0.1466014 DRQSPREP    0.01
PAD440   0.1273735   PAD440    0.01
BPQ020   0.1253917   BPQ020    0.01
RIAGENDR 0.1169647 RIAGENDR    0.01
DBQ095Z  0.1115717  DBQ095Z    0.01
\end{CodeOutput}
\end{CodeChunk}

\subsection{Clustering with minimal forest} \label{dim_red}
\begin{figure}
    \centering
    \includegraphics[scale=.8]{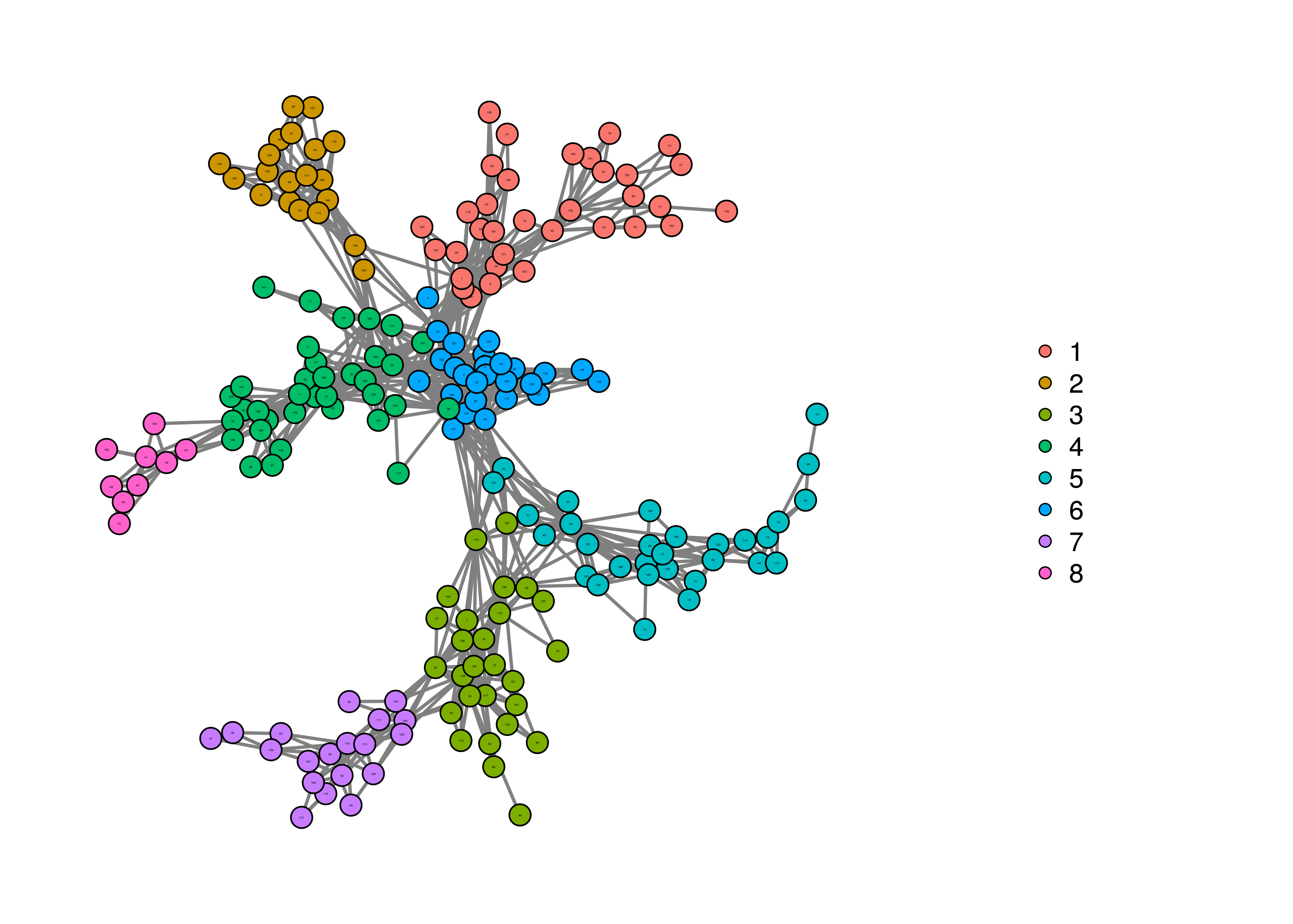}
    \caption{\textbf{Population stratification (clustering) with minimal forest}. The graph is constructed using \code{min.forest} function with \code{BIC} method on the transposed dataset. The nodes represent samples and are colored according to the graph communities.}
    \label{fig:cluster}
\end{figure}

In order to show the functionality of minimal forest in clustering, we select a subsample of size 200 and apply \code{min_forest} function on the subsampled dataset. Therefore the transposed dataset is passed to the funciton. Fig. \ref{fig:cluster} illustrates the stratified population of the samples.\\
\begin{Schunk}
\begin{Sinput}
> t_nhanes <- as.data.frame(sapply(as.data.frame(t(nhanes[1:200, ])), 
function(x) as.numeric(as.character(x))))
> clusters_mf <- min_forest(t_nhanes)
> clusters_vis <- graph_vis(clusters_mf$graph, plot = T, 
filename = "clusters", filetype = "png")
\end{Sinput}
\end{Schunk}

\begin{figure}
    \centering
         \includegraphics[width=.9\textwidth]{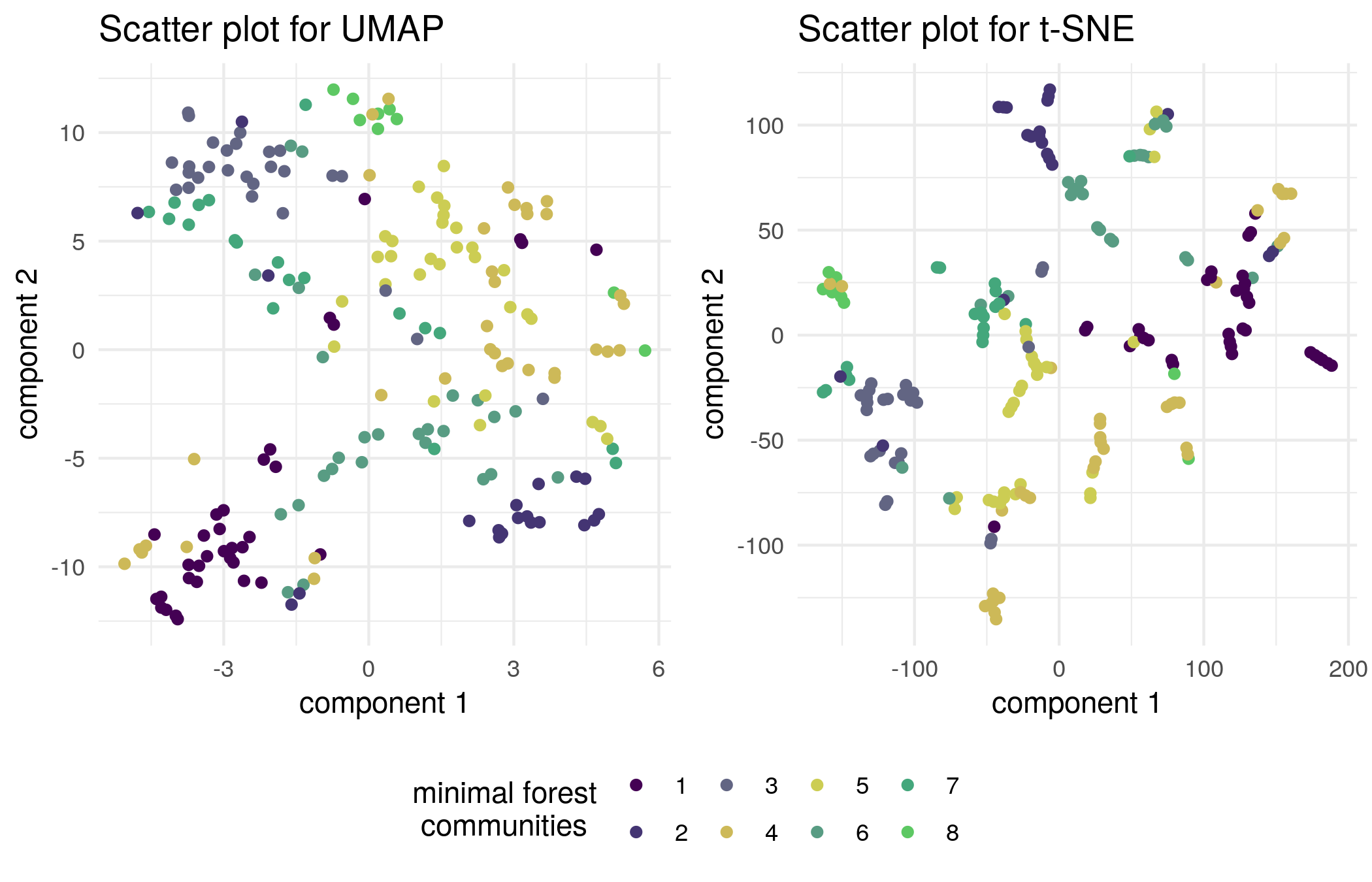}
\caption{\textbf{Two-dimensional visualization of the subsampled dataset}. The funciton \code{dim\_reduce} is used with parameter \code{method} set to \code{umap} (on left) and \code{tsne} (on right) to perform dimensionality reduction. The nodes are colored based on their community in the minimal forest illustrated in Fig. \ref{fig:cluster}.}
\label{fig:dim_red}
\end{figure}

Dimensionality reduction algorithms can be used in order to get a proper visualization of data in low (i.e., two or three) dimensions. We use \code{dim_reduce} function with \code{tsne} and \code{umap} methods to plot the aforementioned subpopulation in two dimensions. The points are colored according to their communities within the minimal forest in Fig. \ref{fig:cluster}. As one can see in Fig. \ref{fig:dim_red} the samples within each community are clustered together, in a fairly accurate manner, particularly with UMAP.
\begin{Schunk}
\begin{Sinput}
> communities <- clusters_mf$communities
> communities <- communities[match(c(1:200), as.integer(names(communities)))]

## Using 'umap' method
> ump <- dim_reduce(nhanes[1:200,], method = "umap", annot1 = as.factor(communities)
, annot1.name = "minimal forest\n communities")

## Using 'tsne' method
> tsn <- dim_reduce(nhanes[1:200,], method = "tsne", annot1 = as.factor(communities)
, annot1.name = "minimal forest\n communities")

## Using cowplot to plot with shared legend
> require(cowplot)
> require(ggplot2)
> leg <- get_legend(ump + theme(legend.position = "bottom"))
> plt <- plot_grid(ump + theme(legend.position = "none"),
tsn + theme(legend.position = "none"))
> plot_grid(plt, leg, ncol = 1, rel_heights = c(1, .2))
\end{Sinput}
\end{Schunk}

\section{Conclusions}
In this document, we introduced \pkg{muvis} as a set of tools for analysis and visualization of multivariate data. The package provides the users with an easy-to-use and end-to-end analysis pipeline. The methods implemented in \pkg{muvis} can be applied on a wide range of datasets with high number of variables and observations. It can be used for data preprocessing (\code{data_preproc}), statistical analysis (\code{VKL}, \code{VVKL}, \code{test_assoc}, \code{ggm}, \code{dgm}, and \code{min_forest}), and visualization (\code{plot_assoc}, \code{graph_vis}, and \code{dim_reduce}). As demonstrated in section \ref{sec_data}, the results assert the functionality of the package on real data: preprocesing method could effectively detect and eliminate the outliers; the GMs were efficient to simultaneously estimate the structure of associations and interpretations; and the visualization method could easily visualize associations among variables. We also introduced novel KL-based methods for determining important variables that could explain surprising observations, including violation of expected linear associations. This work can be extended in several directions: providing predictive models (e.g. elastic nets, ensemble methods, etc.), non-Gaussian GMs, and enhanced imputation methods. 
\section{License}
This package is available under GNU General Public License (GPL) version 3.
\newpage

\begin{table}[H]
\caption {Description of mentioned variables.}
\label{vars}
\centering
\begin{tabular}{|c|c|}
\hline
\textbf{Variable} & \textbf{Label} \\ \hline
SEQN & Respondent sequence number \\ \hline
LBDHDD & Direct HDL-Cholesterol (mg/dL)  \\ \hline
DRQSDIET &  On special diet?\\ \hline
DRQSPREP & Salt used in preparation? \\ \hline
LBXVIC & Vitamin C (mg/dL) \\ \hline
LBXALC &  Alpha-carotene (ug/dL)\\ \hline
DRD360 &  Fish eaten during past 30 days\\ \hline
LBXVIE &  Vitamin E (ug/dL)\\ \hline
PAD590 &  \# hours watch TV or videos past 30 days\\ \hline
DSD010 &  Any Dietary Supplements Taken?\\ \hline
LBXFOL &  Folate, serum (ng/mL)\\ \hline
PAD440 &  Muscle strengthening activities\\ \hline
PAQ100 &  Tasks around home/yard past 30 days\\ \hline
PAD200 &  Vigorous activity over past 30 days\\ \hline
HUQ050 &  \# times receive healthcare over past year\\ \hline
LBXSATSI & Alanine aminotransferase ALT (U/L) \\ \hline
LBXSAL & Albumin (g/dL)\\ \hline
PAQ180 & Avg level of physical activity each day \\ \hline
LBXSCH & Cholesterol (mg/dL) \\ \hline
SMD410 & Does anyone smoke in home? \\ \hline
LBXHCT & Hematocrit (\%) \\ \hline
LBXHGB & Hemoglobin (g/dL) \\ \hline
LBXSIR & Iron, refigerated (ug/dL) \\ \hline
LBXMOPCT &  Monocyte percent (\%)\\ \hline
LBXRBCSI &  Red blood cell count (million cells/uL)\\ \hline
LBXRDW & Red cell distribution width (\%) \\ \hline
LBXSTR & Triglycerides (mg/dL) \\ \hline
LBXSUA & Uric acid (mg/dL) \\ \hline
LBXWBCSI & White blood cell count (1000 cells/uL) \\ \hline
BMXBMI & Body Mass Index (kg/m**2) \\ \hline
LBXBCD & Cadmium (ug/L) \\ \hline
LBXCOT & Cotinine (ng/mL) \\ \hline
BPQ020 & Ever told you had high blood pressure \\ \hline
HUQ010 & General health condition \\ \hline
LBXTHG & Mercury, total (ug/L) \\ \hline
LBXTC & Total Cholesterol (mg/dL) \\ \hline
RIDRETH1 & Race/Ethnicity \\ \hline
RIAGENDR & Gender \\ \hline
PAQ520 & Compare activity w/others same age \\ \hline
HSD010 & General health condition \\ \hline
DBQ095Z & Type of table salt used \\ \hline
\end{tabular}
\end{table}

\bibliography{references}
\newpage
\end{document}